\documentclass[letterpaper,twocolumn,10pt]{article}
\usepackage{usenix2019_v3}
\usepackage{amsmath}
\usepackage{adjustbox}
\usepackage{graphicx}
\graphicspath{{figures/},{pics/}}
\usepackage{xspace}
\usepackage{graphicx}
\usepackage{textcomp}
\usepackage{xcolor}
\usepackage{multirow}
\usepackage{tcolorbox}
\usepackage{booktabs}
\usepackage{colortbl} 
\usepackage{url}
\usepackage{subcaption}
\usepackage{tikz}
\usepackage{amsmath}
\usepackage{algorithm}
\usepackage[noend]{algpseudocode} 
\usepackage{filecontents}
\usepackage{algorithm}
\usepackage{algpseudocode}
\usepackage{amsmath}
\usepackage{hyperref}
\usepackage{placeins}
\usepackage{tabularx}

\begin{document}

\algnewcommand\And{\textbf{and}}
\algnewcommand\Or{\textbf{or}}
\newcommand{\vars}{\texttt}
\date{}

\title{\Large \bf {DeLog}: An Efficient Log Compression Framework with Pattern Signature Synthesis}

\date{%
\footnotetext{This paper uses the Loghub 1.0 and Loghub 2.0 datasets. We obtained consent from the dataset authors and strictly comply with their non-commercial licenses. This work is for research purposes only and has not been used in any ByteDance product or commercial deployment.}%
}
\author{
{\rm Siyu Yu}\textsuperscript{\textdagger}\quad
{\rm Yifan Wu}\textsuperscript{\textdagger}\quad
{\rm Junjielong Xu}\textsuperscript{\textdaggerdbl}\quad
{\rm Ying Fu}\textsuperscript{\S}\quad
{\rm Ning Wang}\textsuperscript{\P}\\
{\rm Maoyin Liu}\textsuperscript{\P}\quad
{\rm Pancheng Jiang}\textsuperscript{\P}\quad
{\rm Xiang Zhang}\textsuperscript{\P}\quad
{\rm Tong Jia}\textsuperscript{\textdagger}\quad
{\rm Pinjia He}\textsuperscript{\textdaggerdbl}\quad
{\rm Ying Li}\textsuperscript{\textdagger}\thanks{Corresponding author: li.ying@pku.edu.cn. This paper uses the Loghub 1.0 and Loghub 2.0 datasets. We obtained consent from the dataset authors and strictly comply with their non-commercial licenses. This work is for research purposes only and has not been used in any ByteDance product or commercial deployment.}
\\[0.6em]
\textsuperscript{\textdagger}Peking University\\
\textsuperscript{\textdaggerdbl}The Chinese University of Hong Kong, Shenzhen\\
\textsuperscript{\S}Southwest Jiaotong University\\
\textsuperscript{\P}ByteDance
}

\maketitle






\begin{abstract}
Parser-based log compression, which separates static templates from dynamic variables, is a promising approach to exploit the unique structure of log data. However, its performance on complex production logs is often unsatisfactory. This performance gap coincides with a known degradation in the accuracy of its core log parsing component on such data, motivating our investigation into a foundational yet unverified question: does higher parsing accuracy necessarily lead to better compression ratio?

To answer this, we conduct the first empirical study quantifying this relationship and find that a higher parsing accuracy does not guarantee a better compression ratio. Instead, our findings reveal that compression ratio is dictated by achieving effective pattern-based grouping and encoding, i.e., the partitioning of tokens into low entropy, highly compressible groups.

Guided by this insight, we design DeLog, a novel log compressor that implements a Pattern Signature Synthesis mechanism to achieve efficient pattern-based grouping. On 16 public and 10 production datasets, DeLog achieves state-of-the-art compression ratio and speed. 
\end{abstract}


\section{Introduction}

Logs record detailed information about various aspects of system states ~\cite{yang2018nanolog,he2021survey}, making them a critical data source for various tasks, such as detecting anomalies~\cite{zhang2019robust}, identifying root causes ~\cite{lu2017log}, diagnosing failures~\cite{zhang2025scalalog}. Most of these tasks need to learn knowledge from historical logs rather than generated real-time logs. For instance, training anomaly detection models typically involves analyzing historical logs to learn patterns of known system anomalies. These logs, often termed "offline logs" by researchers~\cite{wei2023loggrep}, are infrequently accessed for searching. As a result, the main objective for their storage management is to achieve the highest possible compression ratio with acceptable performance for compression and decompression. This necessity has led many companies to adopt policies mandating long-term log retention. For instance, AliCloud enforces a policy requiring logs to be retained for up to 180 days \cite{wei2021feasibility}. At ByteDance, log retention periods vary across scenarios.

\begin{figure}[t]
	\centering
		\includegraphics[width=\columnwidth]{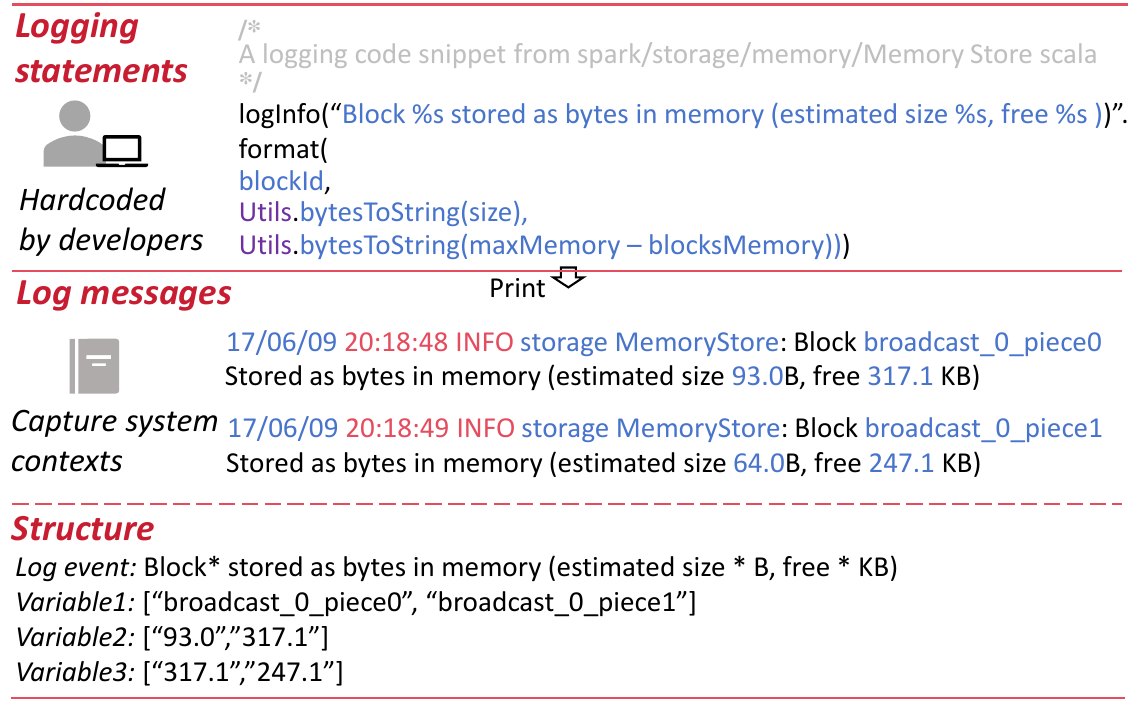}
	   \caption{Log Structure: High redundancy stems from the repetition of static templates, while dynamic variables introduce variability.}\label{intro_structure}
\end{figure}

Storing massive volumes of logs for extended periods incurs high storage costs. Volcano Engine, our collaborator and the cloud platform service provider under ByteDance, generates hundreds of petabytes of logs per day. Although storage costs have shown a declining trend over time, the rapid increase in log data continues to create considerable overhead for service providers ~\cite{wang2024muslope}.
A straightforward approach to reduce storage overhead is compressing log files using general-purpose compressors such as lzma~\cite{lzmalink}, gzip~\cite{gziplink}, and brotil~\cite{brotil}. However, these general-purpose compressors operate on raw byte streams and are oblivious to the intrinsic structure of logs. As illustrated in Figure~\ref{intro_structure}, a log message is composed of a static, repetitive template and a set of dynamic variables. The primary source of redundancy lies in the vast repetition of these identical templates across countless log entries. Because general-purpose compressors cannot distinguish between these two components, they fail to effectively exploit this template-based redundancy~\cite{liu2019logzip,chen2025tracezip}.

To exploit this structural redundancy, researchers introduced \textbf{parser-based compressors}~\cite{wei2021feasibility,wei2023loggrep,rodrigues2021clp,li2024logshrink}. This approach leverages log parsing~\cite{zhu2019tools,he2021survey} to deconstruct each raw message into two distinct components: a static, repetitive template and a set of dynamic variables. By isolating the highly redundant templates, these methods can compress them far more effectively than general-purpose algorithms, promising substantial performance gains.

However, the effectiveness of these parser-based compressors degrades largely on modern production logs, which are characterized by longer entries, a higher density of variables, and mixed data formats. For instance, while a state-of-the-art compressor like LogReducer achieves a 1.3$\times$ to 5.3$\times$ compression ratio improvement over LZMA on public benchmarks, this advantage shrinks to a mere 1.1$\times$ on our production data from Volcano Engine. In diagnosing this dramatic performance gap, we noted that recent studies have consistently reported a sharp degradation in the accuracy of their core component, log parsing, when applied to such complex logs~\cite{zhang2025logbase, xu2023hue}. Intuitively, a drop in parsing accuracy suggests that template redundancy is not being removed as effectively, which could in turn harm the final compression ratio. While this link seems plausible, it has never been systematically verified. This led us to ask a foundational, yet unverified, question: does higher parsing accuracy necessarily lead to a better compression ratio?

To answer this definitively, we first conducted a comprehensive empirical study (Section~\ref{empirical_study}). Our analysis reveals a counter-intuitive result: a higher log parsing accuracy does not necessarily translate to a better log compression ratio. Instead, we discovered that the key to high compression lies not in the pursuit of perfect parsing, but in achieving optimal pattern-based grouping and encoding, the partitioning of tokens into low-entropy, highly compressible streams and encoding them in the appropriate way. We found that such streams are formed when their constituent tokens share common compressible properties, such as value locality (e.g., being drawn from a small, repetitive set of values) or structural predictability (e.g., following a mathematical relationship like monotonic increase). The central challenge, therefore, shifts away from parsing accuracy and towards a new goal: designing an algorithm that can reliably identify these compressible properties for each token.

DeLog is engineered precisely to meet this challenge. It employs a novel, single-pass mechanism to identify a token's compressible properties by synthesizing information from two key sources: its intrinsic structure and external contexts (Section~\ref{feature_pool}). Based on this synthesis, DeLog classifies each token and generates a pattern signature (Section~\ref{tag_gene}). All tokens sharing the same signature are then streamed into the same homogeneous group for tailored encoding. Furthermore, to prioritize user experience, we designed DeLog-L, a variant engineered for maximum decompression speed. By forgoing time-consuming steps like regular expression matching for IP and timestamps during compression, it exchanges a marginal sacrifice in compression ratio for an over 200\% increase in decompression throughput.

Our extensive evaluation on 16 public and 10 production datasets validates our approach. On the 16 public benchmarks, DeLog establishes a new state-of-the-art, achieving a compression ratio \textbf{1.09$\times$ to 1.45$\times$ higher} than baselines, while simultaneously delivering a compression speed that is \textbf{1.87$\times$ to 21.52$\times$ faster}. The effectiveness of this design is confirmed on the 10 real-world production logs from ByteDance \footnote{For privacy reasons, all production logs reported in this paper are anonymized and denoted as 'Log' with a capital letter.} , where DeLog improves the compression ratio by \textbf{1.13$\times$ - 1.38$\times$} over baselines while maintaining \textbf{1.72$\times$ - 28.84$\times$} faster compression speed. Furthermore, its lightweight variant, DeLog-L, achieves a comparable compression ratio and speed while delivering a decompression speed that is \textbf{1.19$\times$ - 292.47$\times$} faster than baselines, dramatically reducing user wait time for data access.

The main contributions of this paper are as follows:

\begin{itemize}
    \item We conduct the first in-depth empirical study on the correlation between log parsing accuracy and compression ratio. Our study reveals a misalignment: higher parsing accuracy does not reliably lead to better compression. We demonstrate that the true driver of compression is optimal pattern-based grouping and encoding, which can outweigh the goal of formal parsing accuracy.

    \item Based on this core insight, we design and implement DeLog, a novel log compression framework. DeLog forgoes traditional, heavyweight parsing in favor of a lightweight, single-pass mechanism that directly partitions log data into homogeneous, pattern-based groups.

    \item We perform an extensive evaluation of DeLog on 16 public and 10 production datasets (more than 100m log lines/100 GB). Our results show that DeLog significantly outperforms state-of-the-art compressors in compression ratio, compression speed, and decompression speed. We also present DeLog-L, a variant that achieves decompression throughput comparable to general-purpose compressors.

    \item We have open-sourced our implementation of DeLog/DeLog-L and the evaluation framework to benefit both researchers and practitioners in the field of log compression~\cite{delog}.
\end{itemize}

\section{Background and Motivation}

\subsection{The Evolving Complexity of Production Logs}

The nature of log data has evolved, challenging traditional log parsing and compression techniques built upon outdated benchmarks. Widely-used datasets like \textit{LogPAI}~\footnote{https://logpai.com}~\cite{he2016experience} no longer reflect the complexity of logs from modern cloud systems, which now contain more complex templates, mixed formats like multi-line stack traces, and feature a template vocabulary orders of magnitude larger than found in benchmarks~\cite{zhang2025logbase, jiang2024large}. As shown in Table~\ref{tab:log_length_comparison}, an average production log entry can be an order of magnitude longer than one from a public dataset. This dramatic increase in length inevitably introduces a higher number of variables and more intricate structural patterns within a single log line. Furthermore, our production logs frequently contain bursts of consecutive variables, a phenomenon rarely seen in the sparse patterns of older datasets. This new reality directly invalidates a core assumption that underpins many state-of-the-art methods: that logs belonging to the same template have the same length. This flawed assumption degrades the effectiveness of length-based grouping heuristics used in popular log parsers like Drain, AEL, and Brain, as well as compressors such as LogZip, LogReducer, and LogShrink.

\begin{table}[htbp]
\centering
\caption{Comparison of Log Entry Length. The table contrasts two randomly selected public benchmarks from LogPAI with 1~m lines from two service components on ByteDance's Volcano Engine.}
\label{tab:log_length_comparison}
\resizebox{0.8\columnwidth}{!}{
\begin{tabular}{@{}llc@{}}
\toprule
\textbf{Log Source} & \textbf{Example Datasets} & \textbf{Avg. Chars / Entry} \\
\midrule
LogPAI Benchmarks & Apache & 98.07 \\
(Public)          & HDFS    & 139.00 \\
     & Linux    & 90,89 \\
          & Spark    & 86.48 \\
\midrule
Our Production Logs & LogA   & 2291.01 \\
(Collected)         & LogB          & 1718.11 \\
        & LogC          & 1418.86 \\
            & LogD         & 1261.74 \\
\bottomrule
\end{tabular}
}
\end{table}

\subsection{Log Compressor}

\begin{figure}[t]
	\centering
		\includegraphics[width=\columnwidth]{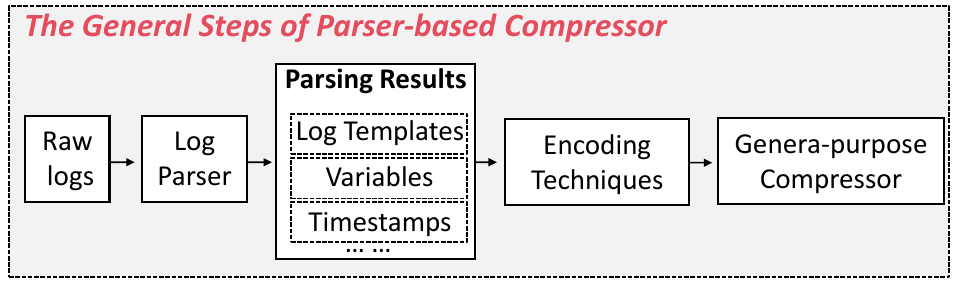}
	  \caption{The General Steps of Parser-based Log Compressor.}\label{parser-based-workflow}
\end{figure}

The general steps of parser-based log compression are illustrated in Figure \ref{parser-based-workflow}. Logs are first processed by a log parser, followed by the application of various encoding techniques. \textbf{Delta Encoding} is widely used to encode numerical content like timestamps, which involves calculating the difference between the current and previous elements for compression. \textbf{Elastic Encoding} is used to compress pure numbers. It uses the most significant bit (MSB) of each byte as a stop bit to achieve flexible encoding of numbers. Upon encountering a byte with an MSB of "\verb|1|", the reading of the current number stops, and the subsequent byte immediately becomes part of the next number, eliminating the need to read four bytes for each number. For instance, the standard integer encoding of "\verb|12|" is "\verb|00000000|, \verb|00000000|, \verb|00000000|, \verb|00001100|," whereas the elastic encoding yields "\verb|10001100|." \textbf{Dictionary Encoding} is a technique to compress highly repetitive content. Specifically, a mapping file is used to store the mapping between elements and their corresponding IDs, the sequences of elements are encoded into ID sequences.

Different parser-based log compressors are mainly distinguished by the encoding techniques they apply to the parsing results. LogZip \cite{liu2019logzip}, the first compressor to integrate a log parser, employs Drain for log parsing and compresses log templates and variables by dictionary encoding. LogReducer \cite{wei2021feasibility} introduces elastic encoding. It also explores potential arithmetic relationships between different types of variables. Building on these advancements, LogShrink \cite{li2024logshrink} further refines the granularity of identifying variables. In addition to parsing-based approaches, there are log compressors that operate without log parsing. However, their compression performance is generally lower \cite{wei2021feasibility}. A recent non-parser-based log compressor is Denum~\cite{yu2024unlocking}, a compressor designed for a specific scenario where logs are predominantly numerical. Denum works by using regular expressions to extract all numerical values and then applies delta encoding to them.

\subsection{The Limitations of Existing Log Compressors on Production Logs}


\begin{figure*}[t]
	\centering
		\includegraphics[width=2\columnwidth]{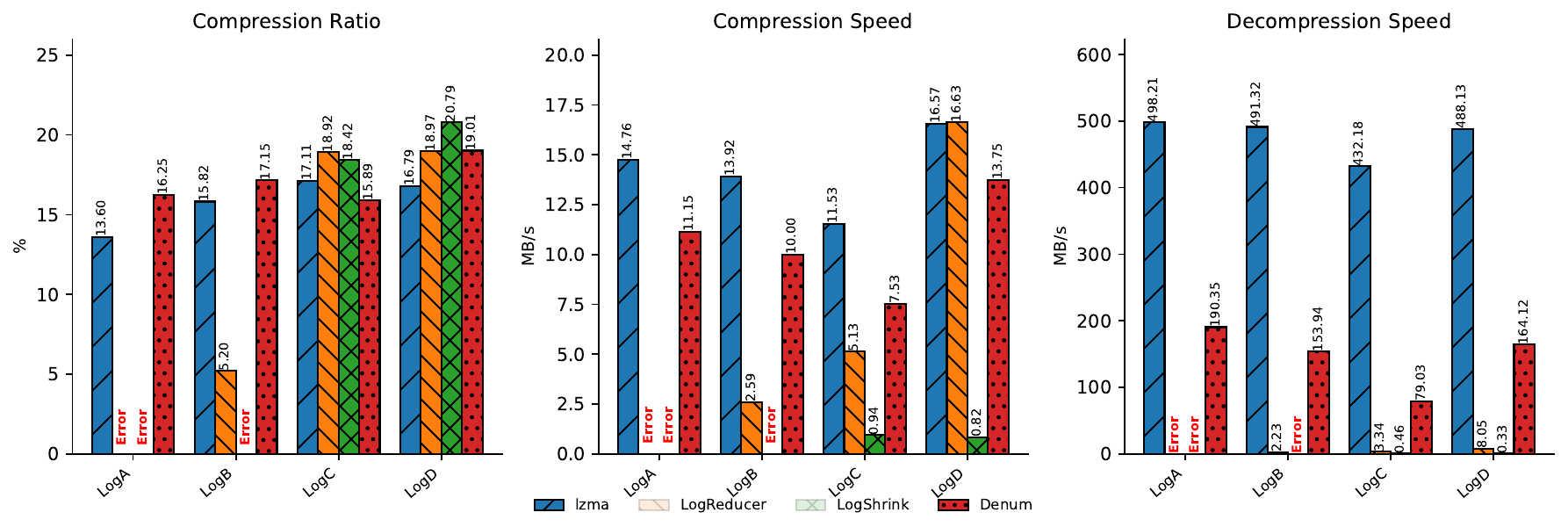}
	  \caption{Performance of Existing Log Compressors on Production Logs: the four log datasets (A-D) are from ByteDance's Volcano Engine, each containing over 1m lines. \textcolor{red}{\texttt{ERROR}} indicates a catastrophic failure (e.g., a segmentation fault) that prevented lossless compression. All compressors run with 4 threads and the highest compression level on 100k-line blocks.}\label{limitations}
\end{figure*}

Existing log compression benchmarks, such as \textit{LogPAI}, are based on decade-old datasets.
Many studies argue that these logs are too simplistic to evaluate how modern parsers and compressors will perform on today's complex production logs.~\cite{jiang2024large,xu2023hue,zhang2025logbase}
When applying state-of-the-art log compressors to production logs from ByteDance, we found their performance to be inadequate as shown in Figure~\ref{limitations}. We identify five key limitations:

\textbf{Poor Compression Ratios.} The compression ratios of existing log compressors drop significantly on production logs compared to public benchmarks. For instance, while LogReducer reportedly achieves a compression ratio 1.30$\times$-5.30$\times$ that of LZMA on public benchmarks, its improvement over LZMA drops to less than 10\% on production logs. Since Denum extracts all numbers and then applies delta compression on them, it generates substantial index overhead when processing logs with numerous complex and irregular numeric tokens. This can even degrade the compression ratio than general-purpose compressors. For instance, on the Log\_C dataset, Denum with an LZMA kernel yields a lower compression ratio than standalone LZMA. 

\textbf{Prohibitive Compression/Decompression Latency.} The compression/decompression speed of these log compressors is unacceptably slow. For example, LogShrink's compression throughput is merely around 1~MB/s, and decompression throughput is around 0.5~MB/S, meaning an SRE~(Site Reliability Engineering) would wait approximately an unacceptable 35 minutes to access a 1~GB log file. While both LogReducer and Denum offer acceptable compression speed, LogReducer's performance is highly asymmetric. Its decompression speed is a major bottleneck, measuring at only ~1/50th that of LZMA and ~1/20th that of Denum.

\textbf{Lack of Robustness and High Compression Overhead.} In a bid to optimize performance, existing log compressors like LogReducer and LogShrink employ a static memory pre-allocation strategy. This design choice necessitates a multitude of hard-coded limits. However, the implementations of these two compressors handle boundary overflows inadequately. Consequently, when processing complex production logs, which frequently exceed these predefined thresholds, we consistently encountered \texttt{Segmentation Fault} errors (due to out-of-bounds array access). Even after attempting to manually patch these limits to prevent crashes, the tool remained inherently fragile, struggling to correctly compress and decompress certain datasets.
Moreover, this pre-allocation strategy introduces a more fundamental issue: exorbitant memory consumption. LogReducer, for instance, exemplifies this problem in its default configuration, compressing a batch of production log required an unsustainable 22~GB of RAM. This resource-intensive behavior fundamentally challenges its practicality and reliability in modern, complex logging environments.

\textbf{Violation of Lossless Compression.} A critical flaw arises when compressors treat numerical strings as mathematical values rather than literal text. To enable optimizations like delta encoding, these tools parse a string like \texttt{09} into its integer equivalent, `9`. While this allows for arithmetic operations (e.g., calculating differences between consecutive values), it inherently discards formatting information such as leading zeros. This process fundamentally violates the principle of lossless compression and can break downstream systems that rely on fixed-width or zero-padded formats. For instance, a timestamp-parsing regex like \texttt{\string\d\{2\} \string\d\{2\}:\string\d\{2\}:\string\d\{2\}} would correctly match the original string \texttt{09 12:23:43} but fail on the decompressed output \texttt{9 12:23:43}.

\subsection{Opportunities}\label{metric_intro}

Existing log compressors present a clear trade-off. Specialized tools like Denum, designed for simple, digit-heavy logs, lack the generality to handle complex production data. While parser-based compressors offer greater generalizability, their performance also degrades substantially on production logs, remaining unsatisfactory.

In diagnosing this performance gap, our attention turned to their log parsing components. Recent studies have consistently reported that the accuracy of log parsers degrades sharply, for instance, Drain's accuracy plummets from 0.867 on benchmarks to just 0.692 on production logs containing mixed formats~\cite{xu2023hue}. Similarly, F1-scores can drop from 0.75 on traditional datasets to as low as 0.55 on more diverse modern benchmarks like LogHub 2.0~\cite{jiang2024large}. This raises a critical question: can a more advanced parser, specifically designed for complex logs, unlock superior compression performance?

Despite the intuitive link between parsing and compression, our literature review revealed a critical gap: no prior work has systematically quantified the relationship between log parsing accuracy and compression performance. To address this, we first conduct a comprehensive empirical study to establish this link.

\section{An Empirical Study on the Correlation between Parsing Accuracy and Compression Ratio}\label{empirical}
\label{empirical_study}

To identify the most promising direction for advancing log compression, we conducted a comprehensive empirical study designed to answer two questions. The full details of this empirical study, including methodology, experimental setup, detailed experimental results, and analysis, are available in the \textbf{Supplemental Material}.

\begin{itemize}
    \item[$\bullet$] Does a higher parsing accuracy guarantee a higher compression ratio?
    \item[$\bullet$] If not parsing accuracy, what properties of a processed log make it highly compressible?
\end{itemize}

\subsection{Methodology and Key Results}

To answer Q1, we first gathered parsing results from six representative log parsers across 12 diverse datasets from LogHub 2.0 \footnote{https://github.com/logpai/loghub-2.0} (over 15m lines). For each combination of parser and dataset, we measured six standard parsing accuracy metrics (i.e., PA, GA, FGA, PTA, RTA, FTA)~\cite{khan2022guidelines} and the resulting compression ratio (CR). Grouping Accuracy (GA) evaluates if logs from the same template are grouped together. Parsing Accuracy (PA) measures if each token in a log is correctly labeled as constant or variable. Template Accuracy (TA) and its variants (PTA, RTA, FTA) assess whether templates are correctly identified and all their corresponding logs are correctly assigned. We then computed the Spearman correlation between each accuracy metric and the final compression ratio. To establish an upper bound, we also included results from a perfect parser using the ground truth.

The aggregated results, summarized in Figure~\ref{spearman_all}, definitively show a weak correlation between parsing accuracy and compression. Across all datasets, the correlation is consistently low. The highest correlation coefficient observed, for Parsing Accuracy (PA), is a mere 0.280, far below the 0.7 threshold typically considered indicative of a strong relationship~\cite{khan2024impact,fu2023empirical}. This result demonstrates that simply improving a parser's formal accuracy is not a reliable path to better compression.

\begin{figure}[h]
    \centering
    \includegraphics[width=\columnwidth]{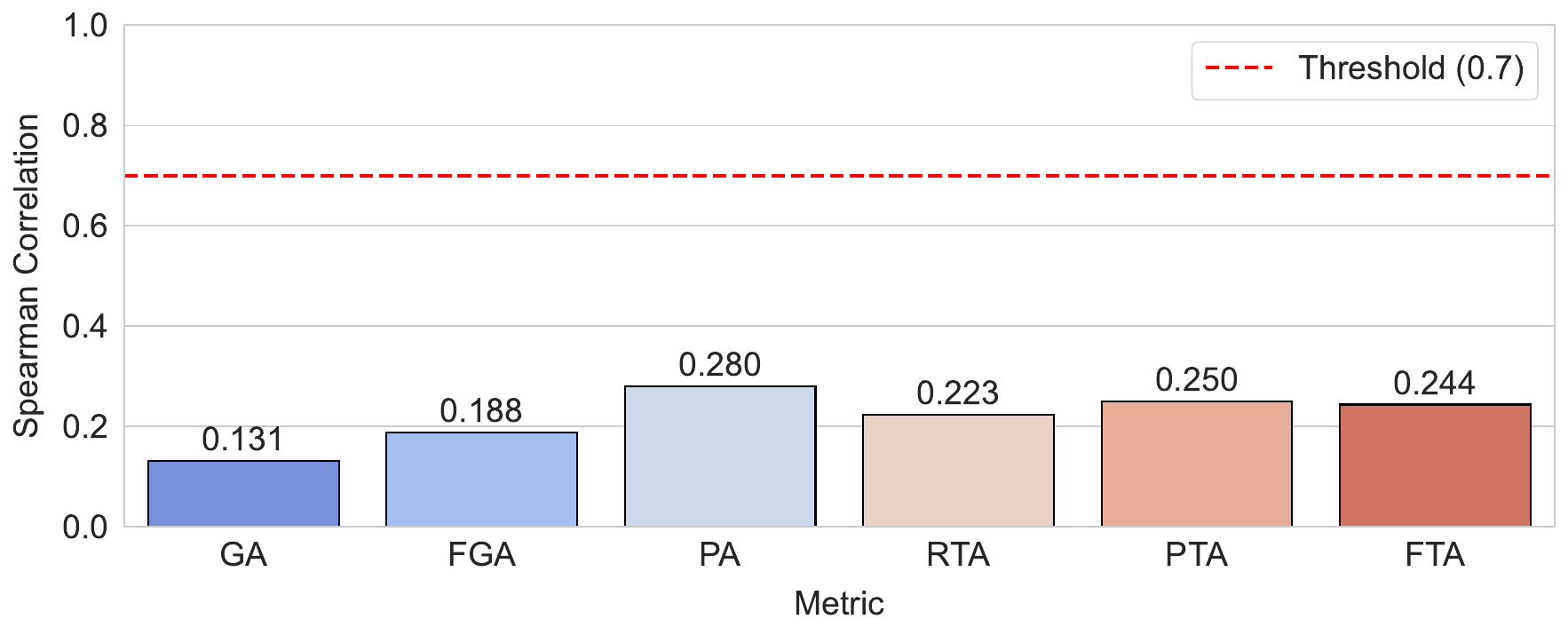}
    \caption{Spearman Correlation Between Log Parsing Accuracy Metrics and CR (Aggregated Across 12 Datasets from Loghub 2.0).}\label{spearman_all}
\end{figure}

\begin{tcolorbox}[colback=gray!10, colframe=black, sharp corners, boxrule=0.8pt]
\textbf{Answer to Q1:} Higher parsing accuracy does not reliably lead to a better compression ratio.
\end{tcolorbox}

The weak correlation found in Q1 compels us to answer Q2. Our detailed analysis reveals that the true driver of high compression is not parsing accuracy itself, but the creation of homogeneous, pattern-based streams of data.

Log parsing, typically comprising two main stages, i.e., formatting and content template extraction, can be viewed as a multi-stage application of this principle. The first stage, formatting, separates structured header fields like timestamps from the unstructured message content. This initial step is a coarse but powerful form of pattern-based grouping because it naturally creates streams with strong compressible properties. For example, the stream of timestamps exhibits high structural predictability, as values typically increase with a small, stable delta. Similarly, the stream of log levels (e.g., \texttt{INFO}, \texttt{ERROR}, \texttt{FATAL}) exhibits high value locality, as it is drawn from a very small and highly repetitive set of strings. Our experimental results confirm that this simple partitioning alone contributes the majority of the compression gains across most benchmarks.

The subsequent step, content template extraction, is a more refined application of grouping. Its effectiveness hinges on its ability to isolate streams with strong compressible properties. This is evident on the Apache dataset, where the \texttt{Child ID} variable forms a stable arithmetic sequence with a delta of 1. A parser that correctly extracts these IDs creates a stream with high structural predictability, achieving a remarkable 35\% CR improvement over simple formatting because the stream is perfectly suited for delta encoding. In contrast, a parser like LogCluster~\cite{makanju2009clustering}, which fails to isolate this predictable stream, sees its CR improve by only 5\%.

This principle is most powerfully illustrated when a formally "incorrect" parse creates groups with superior compressible properties. On the Proxifier dataset, the ground truth parser correctly groups all domain names into a single stream, which has low value locality due to the mix of internal and external domains. In contrast, the parser IPLoM "incorrectly" splits them into two more refined groups: one for LAN domains and one for remote domains. This was superior for compression because each of these new streams exhibits much higher value locality, LAN domains are more likely to repeat among themselves, as are remote domains. Consequently, each stream is far more compressible via dictionary encoding than the single, mixed stream from the ground truth. This leads to our most critical insight: the ultimate goal is to achieve pattern-based grouping, even if it means sacrificing formal parsing accuracy.

\begin{tcolorbox}[colback=gray!10, colframe=black, sharp corners, boxrule=0.8pt]
\textbf{Answer to Q2:} The key property for high compression is pattern homogeneity in the resulting data streams. We find that the compression gains from log parsing are not a direct result of its accuracy, but rather because parsing itself implicitly performs a degree of pattern-based grouping. 
\end{tcolorbox}

\subsection{From Findings to Design: Motivating DeLog}
\label{feature_two}

Our empirical findings dictate a new design principle: to maximize log compression, a pre-processor must prioritize creating pattern-homogeneous groups over achieving formal parsing accuracy. This raises the central design question: how can such patterns/compressible properties be identified effectively? Our analysis on a huge amount of logs reveals that a robust pattern identification mechanism cannot rely on a single source of information. Instead, it must synthesize information from distinct sources:

\begin{itemize}
    \item \textbf{Intrinsic Structure:} The pattern is self-evident from the token's intrinsic structure. A token like \texttt{19:23:23,456} is immediately recognizable as a timestamp due to its combination of digits and specific special characters (\texttt{:,}). A stream of such timestamps, for instance, exhibit strong structural predictability and will typically increase with a small and relatively stable delta. By grouping these tokens based on their shared intrinsic structure, we create a stream of underlying numeric values that is highly amenable to compression techniques like delta encoding.
    
    \item \textbf{External Context:} The pattern is defined by surrounding keywords, especially when the token itself is semantically ambiguous. This principle stems from the fact that logs are fundamentally designed for human interpretation. When instrumenting code, developers intentionally add semantically rich keywords (e.g., \texttt{Temperature=}, \texttt{UserID=}) to provide context for variables that lack distinct intrinsic structure. For instance, the number \texttt{78} is generic on its own, but the preceding keyword \texttt{Temperature=} immediately and reliably clarifies its pattern as a temperature reading. This context-based grouping is powerful because it creates a data stream with excellent compressible properties. A stream of machine room temperatures, for example, typically stabilizes within a narrow range (e.g., 10-20°C), creating a group with high \textbf{value locality} that is highly amenable to dictionary or elastic encoding.
    
    \item \textbf{Combined Structure and Context:} The pattern requires synthesizing both intrinsic and external cues. In the logs \texttt{from IP 23.4.1.102} and \texttt{from IP 2001:db8::...}, the external context (\texttt{from IP}) is identical. Only by also analyzing the token's intrinsic structure can we disambiguate between an IPv4 and an IPv6 address pattern, thus creating two separate, highly homogeneous groups with low value locality.
\end{itemize}

Considering multi-faceted features as mentioned above to achieve pattern-based grouping form the core design philosophy of our proposed system, DeLog. The subsequent section details the design of DeLog.

\begin{figure*}[t]
	\centering
		\includegraphics[width=\textwidth]{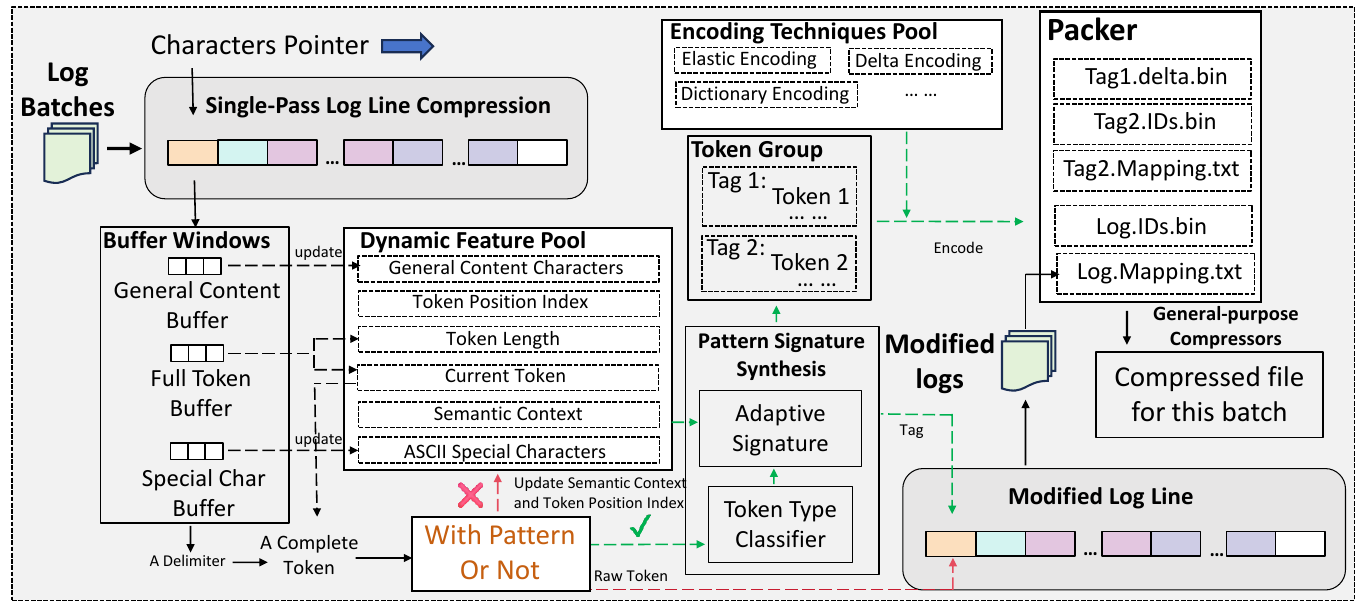}
	  \caption{The Overview of \textit{DeLog}.}\label{overview}
\end{figure*}

\section{Our Approach}

Inspired by these key findings, we have built DeLog, a novel log compression framework with about 3,000
 lines of C/C++ code. Our approach introduces a stateful, single-pass word signature synthesis engine that constructs pattern-homogeneous groups. Figure~\ref{overview} illustrates the overview of DeLog, which comprises four main modules: 1. \textbf{Dynamic Feature Pool Construction.} DeLog performs tagging in a single pass over the log data. As it scans the log character-by-character, it captures both intrinsic structure and external context, storing them in a lightweight Feature Pool. As soon as a complete token is identified by the single-pass scanner, its corresponding Feature Pool is finalized.
2. \textbf{Pattern Signature Synthesis.} The feature pool is then immediately passed to the classification module, which performs two functions. First, it classifies the token into a predefined category (e.g., keyword). Second, based on this category, it adaptively selects the most relevant features from the pool to generate a structured signature. All tokens that are ultimately assigned the same signature are partitioned into the same group. These groups are then compressed using different encoding techniques tailored to their respective categories.
3. \textbf{Final-stage Compression.} Finally, all processed files, including the encoded token groups and the compressed templates, are bundled into a single archive. This archive is then compressed by a general-purpose algorithm (e.g., Gzip, LZMA) to generate the final compressed file. 4. \textbf{Decompression process of DeLog.} a reverse process of compression.

\begin{table}[htbp]
\centering
\caption{Structure of the Dynamic Feature Pool.}
\label{tab:feature-pool}
\resizebox{0.8\columnwidth}{!}{
\begin{tabularx}{\columnwidth}{@{} l >{\raggedright\arraybackslash}X @{}}
\toprule
\textbf{Feature} & \textbf{Details} \\ \midrule
\vars{current\_token} & \textbf{Type}: String, \textbf{Source}: Intrinsic Structure. The original, unmodified token string. \\ \addlinespace

\vars{token\_length} & \textbf{Type}: Integer, \textbf{Source}: Intrinsic Structure. The character length of the token. \\ \addlinespace

\vars{ascii\_special} & \textbf{Type}: String, \textbf{Source}: Intrinsic Structure. A concatenation of all ASCII special characters within the token. \\ \addlinespace

\vars{general\_content} & \textbf{Type}: String, \textbf{Source}: Intrinsic Structure. All non-ASCII-special characters, including letters, digits, and multi-byte characters. \\ \addlinespace

\vars{semantic\_context} & \textbf{Type}: String, \textbf{Source}: External Context. The most recently observed keyword token. \\ \addlinespace

\vars{token\_index} & \textbf{Type}: Integer, \textbf{Source}: External Context. The positional index of a token within a sequence of non-keywords. \\ \bottomrule
\end{tabularx}
}
\end{table}

\begin{algorithm}[t]
\caption{Dynamic Feature Pool Construction}
\label{alg:feature-population}
\begin{algorithmic}[1]
\Procedure{ProcessLine}{\vars{logLine}}
    \State $\vars{ModifiedLog}, \vars{context} \gets \text{empty string}$
    \State $\vars{tokenIndex}, \vars{ptr} \gets 0$ 
    \While{$\vars{ptr} < \text{length}(\vars{logLine})$}
        \State \textproc{AppendDelimiters}(\vars{logLine}, \vars{ptr}, \vars{ModifiedLog})
        
        \If{$\vars{ptr} < \text{length}(\vars{logLine})$}
            \State Initialize buffers: $\vars{token}, \vars{content}, \vars{speciall}, \vars{len} \gets \text{""}, \text{""}, \text{""}, 0$
            \While{$\vars{ptr} < \text{length}(\vars{logLine}) \And \neg \text{IsDelimiter}(\vars{logLine}[\vars{ptr}])$}
                \State $c \gets \vars{logLine}[\vars{ptr}]$
                \State Append $c$ to $\vars{token}$; $\vars{len} \gets \vars{len} + 1$
                \If{IsAsciiSpecial($c$)} \State Append $c$ to $\vars{special}$
                \Else \State Append $c$ to $\vars{content}$
                \EndIf
                \State $\vars{ptr} \gets \vars{ptr} + 1$
            \EndWhile

            \State $\vars{pool} \gets \text{new FeaturePool()}$
            \State $\vars{pool.fields} \gets \{\vars{token}, \vars{len}, \vars{content}, \vars{special}, \vars{context}, \vars{tokenIndex}\}$

            \State $(\vars{containPattern}, \vars{replacement}) \gets \textproc{Signature}(\vars{pool})$
            \State Append $\vars{replacement}$ to $\vars{ModifiedLogs}$
            
            \If{$\vars{containPattern}$} \State $\vars{tokenIndex} \gets \vars{tokenIndex} + 1$
            \Else
                \State $\vars{context} \gets \vars{token}$
                \State $\vars{tokenIndex} \gets 0$
            \EndIf
        \EndIf
    \EndWhile
    \State \Return $\vars{ModifiedLogs}$
\EndProcedure
\end{algorithmic}
\end{algorithm}

\subsection{Dynamic Feature Pool Construction}
\label{feature_pool}
Algorithm~\ref{alg:feature-population} details the dynamic construction of our feature pool. DeLog performs a single-pass scan over the raw log line, character by character, to build a comprehensive feature pool for each potential token. This efficient approach eliminates any need for secondary processing passes. The resulting Feature Pool, formally defined in Table~\ref{tab:feature-pool}, aggregates two distinct classes of features, that is, \textit{intrinsic structure} and \textit{external context} (Section~\ref{feature_two}), which are then used for pattern signature synthesis.

The \textit{intrinsic structures}, derived from the token's content, are collected on the fly using concurrent buffers. As the algorithm scans a token's characters (Line 8), it simultaneously populates multiple buffers: a \texttt{full\_token} buffer for the complete, unmodified string; a \texttt{general\_content} buffer containing all non-ASCII-special characters (including letters, digits, and multi-byte characters like Chinese); and an \texttt{ascii\_special} buffer that isolates all special characters (Lines 10–13). This provides a multi-faceted view of the token's intrinsic structure. Once the token is fully identified, the contents of these buffers are used to populate the intrinsic portion of the feature pool, enabling a comprehensive analysis without secondary processing.

The \textit{external contexts} capture the token's semantic context, which DeLog intelligently updates after each token is processed. If the token is classified as a static keyword, its value becomes the new context for all subsequent tokens, and a positional variable index is reset (Lines 23-25). If the token is a variable, the context remains unchanged, and only the index is incremented (Line 22). This stateful propagation is crucial. For instance, in a log like \texttt{node 0123 0124 0125 joined}, the context remains stably set to \texttt{node} for all three numbers, preventing a "signature explosion" where the context for \texttt{0124} would incorrectly become \texttt{0123}.

Finally, once a token is fully identified, all collected features are synthesized to populate a new \texttt{FeaturePool} instance (Line 16-17). This complete feature set is then passed to the next module for pattern signature (Line 18).

\begin{table}[t]
\centering
\caption{Pattern Signature Synthesis Rules.}
\label{tab:tagging-rules}
\resizebox{0.8\columnwidth}{!}{
\begin{tabularx}{\columnwidth}{@{} l >{\raggedright\arraybackslash}X @{}}
\toprule
\textbf{Token Pattern} & \textbf{Details} \\ \midrule
\textbf{Keyword (No pattern)} & 
    \textbf{Condition}: Contains no digits and no ASCII special characters. \newline
    \textbf{Example}: \texttt{node}, \texttt{Starting}, \texttt{Error} \newline
    \textbf{Signature}: \textit{(Original token is kept)} \\
\addlinespace

\textbf{Short Pure Numeric} & 
    \textbf{Condition}: Purely numeric and length $\le 2$. \newline
    \textbf{Example}: \texttt{9}, \texttt{01} \newline
    \textbf{Signature}: \texttt{<LEN=length>} \\
\addlinespace

\textbf{Long Pure Numeric} & 
    \textbf{Condition}: Purely numeric and length $> 2$. \newline
    \textbf{Example}: \texttt{12345}, \texttt{20240101} \newline
    \textbf{Signature}: \texttt{<IDX=idx|CTX=ctx|LEN=length>} \\
\addlinespace

\textbf{Complex Numeric} & 
    \textbf{Condition}: Contains digits and special characters (no letters). \newline
    \textbf{Example}: \texttt{1.2.3}, \texttt{12-34} \newline
    \textbf{Signature}: \texttt{<CTX=ctx|STR=pat>} \\
\addlinespace

\textbf{Alphanumeric} & 
    \textbf{Condition}: Contains letters along with digits and/or special characters. \newline
    \textbf{Example}: \texttt{node\_1}, \texttt{dev-abc} \newline
    \textbf{Signature}: \texttt{<CTX=ctx|STR=\_chars>} \\
\bottomrule
\end{tabularx}
}

\vspace{0.2em} 
\footnotesize{\textit{Note:} \texttt{ctx} is the last keyword token; \texttt{idx} is the positional index; \texttt{pat} is a regex-like pattern (e.g., \texttt{\string\d\{3\}.\string\d\{1\}.\string\d\{1\}.\string\d\{1\}}); and \texttt{\_chars} represents `\_` prepended to special characters (e.g., \texttt{\_--}).}
\end{table}

\subsection{Pattern Signature Synthesis.}
\label{tag_gene}

Unlike traditional log parsers that perform a simple binary classification into "template" and "variable," DeLog employs a more sophisticated approach to create pattern-based groups. First, upon identifying a token, it performs a preliminary pattern check. If the token is a simple Keyword (containing only natural language text), it is considered to have no pattern/compressible properties. These tokens form the static backbone of the modified log and are kept in their original form without generating a pattern signature. 
All other tokens are deemed to have a pattern/compressible properties and proceed to the further pattern signature synthesis stage. In this stage, DeLog classifies the token into a more specific pattern category, such as \texttt{Short Pure Numeric} or \texttt{Complex Numeric}, and then adaptively selects relevant features from the Dynamic Feature Pool to synthesize a structured, descriptive signature. The specific rules for this multi-tiered classification and the corresponding signature generated for each pattern are detailed in Table~\ref{tab:tagging-rules}.

The design of our classification rules is guided by our goal of creating groups that exhibit homogeneous pattern and compressible properties, i.e., either strong structural predictability or high value locality.
To exploit structural predictability, we were inspired by tools like Denum that reduce the magnitude of numeric values. We created the Long Pure Numeric category specifically to isolate streams of sequential data, such as timestamps, which are ideal candidates for delta encoding. This strategy is not applied to Short Pure Numeric tokens, as their small integer values do not benefit from this type of arithmetic reduction.
To enhance value locality, we employ two primary features. The most powerful is the semantic context (\texttt{CTX}), which is applied to all categories except Short Pure Numeric. Context is crucial for disambiguating tokens that may be structurally similar but semantically distinct (e.g., separating \texttt{source\_ip} from \texttt{dest\_ip}). We exempt Short Pure Numeric tokens because they are drawn from a very small set of values and thus inherently possess high value locality on their own. As a powerful secondary heuristic, we also incorporate special characters (\texttt{STR}) into the pattern signature generation for any token that contains them. These characters often signal a token's type with high confidence, providing a strong basis for grouping and reinforcing the distinctions made by the semantic context.
The specific rules for this classification, which synthesize these features to generate a final pattern signature for each pattern, are detailed in Table~\ref{tab:tagging-rules}.

To eliminate repetitive complex pattern signatures, the full descriptive signature generated for a token is not directly inserted. Instead, it serves as a key in a global pattern signature manager, which maps it to a single, compact identifier (e.g., \texttt{<T01>}). This transforms the original log line into a highly regular modified log (e.g., `\texttt{... source\_ip= <T05> dest\_ip= <T06>}`), while the original token values are partitioned into their respective pattern-homogeneous groups.

\subsection{Encoding and Final-stage Compression}

In the final stage, the values within these partitioned groups are efficiently encoded. Our system applies a tailored strategy for each group based on its pattern. For instance, the 'Long Pure Numeric' group is compressed using a combination of delta and elastic encoding, while other groups like 'Alphanumeric' primarily use dictionary encoding. This hybrid approach ensures each data type is compressed with the method best suited to its characteristics, maximizing the overall compression ratio.

Token groups sharing the same encoding method are merged into a single large file. For instance, for all dictionary-encoded token groups, it concatenates the keys into one block and the values into another. An index file is created to record the offset and length of each original data block within the merged file. This approach is critical for performance. In complex log systems with potentially tens of thousands of variable types, managing numerous small files would cause excessive I/O and severely degrade both compression and decompression speed. Finally, all consolidated data and index files are bundled into a single archive. This archive is then compressed using a pluggable general-purpose compressor, such as LZMA, gzip, or bzip2.

\textbf{Delog-L.} To achieve optimal decompression speed, the DeLog-L variant omits all regular expression matching for common variables like IP and Timestamps. This omission constitutes the sole architectural difference between DeLog-L and the standard DeLog. The rationale is that regex matching is inherently CPU-intensive. Its complex, stateful logic operates character-by-character, leading to frequent branch mispredictions and pipeline stalls that cripple modern processors' performance. By replacing regex with direct, fixed-offset parsing, DeLog-L eliminates this bottleneck, significantly reducing the number of CPU cycles required per decompressed byte.

\subsection{Decompression Process of DeLog}

DeLog's decompression is a lossless process that precisely reverses the compression steps, operating in parallel on data chunks. Initially, the compressed archive for each chunk is unpacked to access its core components: the sequence of modified logs, the pattern signature mappings, and the streams of encoded values. The decompressor then reconstructs each original log line by processing the modified logs, applying the appropriate decoding routine (e.g., delta-decoding for numeric sequences, or dictionary lookup) for each pattern signature placeholder to retrieve its value. A critical step is the lossless reconstruction of the original token using its descriptive pattern signature. This is particularly important for numbers. For example, if a signature's pattern is \texttt{\string\d\{4\}-\string\d\{2\}} and the stored numeric value is \texttt{20241}, the system uses the \texttt{\string\d\{2\}} constraint to restore the token as \texttt{2024-01}, automatically re-inserting the leading zero. This mechanism guarantees that all tokens are restored byte-for-byte to their original form, solving the common issue of data loss when treating numbers as values.

\section{Evaluation}

We conduct comprehensive evaluation experiments by applying DeLog on the most widely-used benchmark, Loghub and 10 production logs with the size of around 100GB and report the results to answer the following questions:
\begin{itemize}
	\item[$\bullet$] What is the performance of DeLog on benchmark?
 	\item[$\bullet$] What is the performance of DeLog on production logs?
 	\item[$\bullet$] How does each design contribute to DeLog?
\end{itemize}

\subsection{Experimental Setting}
\textbf{Dataset. }Our experiments employ the widely accepted benchmark datasets sourced from Loghub \cite{he2020loghub}. The dataset comprises over 378 million log entries, totaling more than 77GB, and is collected from 16 diverse systems. These systems encompass distributed system logs like HDFS and OpenStack, supercomputing logs such as BGL, operating system logs like Linux, and server applications including Apache and OpenSSH. In addition to these public benchmarks, we collected 10 large-scale production log datasets from diverse services at ByteDance to evaluate performance under realistic, modern conditions. These datasets, anonymized as Log A through Log J, are summarized in Table~\ref{tab:production_logs}.


\begin{table}[h]
\centering
\caption{Statistics of Production Log Datasets from ByteDance.}
\label{tab:production_logs}
\resizebox{0.5\columnwidth}{!}{
\begin{tabular}{@{}lcr@{}}
\toprule
\textbf{Dataset} & \textbf{Size} & \multicolumn{1}{c}{\textbf{Messages}} \\ \midrule
Log A & 38.30 GB & 164,560,671 \\
Log B & 11.72 GB & 100,000,000 \\
Log C & 5.77 GB & 101,182,412 \\
Log D & 7.59 GB & 47,387,603 \\
Log E & 4.11 GB & 17,333,634 \\
Log F & 3.62 GB & 61,459,876 \\
Log G & 0.23 GB & 3,178,232 \\
Log H & 5.09 GB & 75,985,203 \\
Log I & 0.12 GB & 1,000,000 \\
Log J & 39.85 GB & 301,069,784 \\
\bottomrule
\end{tabular}
}
\end{table}

\textbf{Evaluation Metrics.} In our experiments, we employed Compression Ratio (CR), Compression Speed (CS) and Decompression Speed (DCS) to evaluate the effectiveness and efficiency of DeLog, respectively. \textcolor{red}{\texttt{ERROR}} in figures and tables indicates a catastrophic failure (e.g., a segmentation fault)




\textbf{Baselines.} We selected the state-of-the-art log compressor, Denum, along with all the log compressors used in its experiments as baselines, excluding LogZip (too slow for complex production logs). While other notable log compressors, such as LogArchive \cite{christensen2013adaptive}, LogBlock \cite{yao2021improving}, and Cowic \cite{lin2015cowic}, exist, they were excluded from our evaluation for specific reasons. LogArchive has been demonstrated to underperform significantly compared to LogReducer \cite{wei2021feasibility}. Meanwhile, Cowic \cite{lin2015cowic} prioritizes decompression speed over CR, and LogBlock is optimized for compressing small log blocks. For metrics that are independent of the experimental setup, such as CRs, we directly utilized the data reported from researchers. For environment-dependent metrics, including CSs, we reimplemented all baselines in our experimental environment to ensure fair and consistent comparisons. Due to page limits, we report the experimental results on the performance of general-purpose compressors over both public benchmarks and production logs in the \textbf{Supplementary Material}.

\textbf{Implementation and Environment.} All experiments are carried out inside a Linux 11.9 container running on Linux kernel 5.4.143.
Our container instance is configured with 16 CPU cores and 32 GB of memory.
The source code is compiled using GCC 10.2.1 under the C++17 standard. For each compressor, we employ 4 threads to parallelize the compression of log data. DeLog leverages the "tar" utility (version 1.30) to generate packer files, with lzma serving as the final general compressor. The regular expression functionality was implemented using PCRE2, with PCRE2\_CODE\_UNIT\_WIDTH configured to 8. Following established configuration in prior research, all input log data is partitioned into blocks of 100K lines.


\begin{table}[ht]
\centering
\caption{The Compression Ratio (CR) of DeLog and Baselines. The best result is in \textbf{bold}.}
\label{tab:compression_ratio}
\resizebox{\columnwidth}{!}{
\begin{tabular}{lccccc}
\toprule
Dataset & LogReducer & LogShrink & Denum & DeLog & DeLog-L \\
\midrule
Android     & 20.776 & 21.857 & \textbf{32.494} & 30.354 & 28.634 \\
Apache      & 43.028 & 55.940 & 58.517 & \textbf{59.648} & 49.872 \\
BGL         & 38.600 & 42.385 & 41.804 & \textbf{45.682} & 36.106 \\
Hadoop      & 52.830 & 60.091 & 78.546 & \textbf{79.626} & 67.660 \\
HDFS        & 22.634 & 27.319 & 25.670 & \textbf{29.013} & 28.386 \\
HealthApp   & 31.694 & 39.072 & 44.472 & \textbf{55.934} & 31.921 \\
HPC         & 32.070 & 35.878 & 45.275 & \textbf{46.842} & 26.946 \\
Linux       & 25.213 & 29.252 & 30.449 & \textbf{30.626} & 21.149 \\
Mac         & 35.251 & 39.860 & 40.789 & \textbf{43.687} & 40.923 \\
OpenSSH     & 86.699 & 103.175 & 101.654 & \textbf{106.012} & 60.313 \\
OpenStack   & 16.701 & 22.157 & 22.238 & \textbf{23.754} & 21.783 \\
Proxifier   & 25.501 & 27.029 & 27.288 & \textbf{30.784} & 28.528 \\
Spark       & 59.470 & 59.739 & 59.470 & \textbf{61.788} & 48.701 \\
Thunderbird & 49.185 & 48.434 & 63.824 & \textbf{68.647} & 61.984 \\
Windows     & 342.975 & 456.301 & 481.350 & \textbf{541.569} & 519.578 \\
Zookeeper   & 94.562 & 116.981 & 135.251 & \textbf{154.926} & 47.373 \\
\midrule
Average     & 60.928 & 74.091 & 80.568 & \textbf{88.056} & 69.991 \\
\bottomrule
\end{tabular}}
\end{table}

\subsection{Q3: The Performance of DeLog on Public Benchmarks}

\textbf{Compression Ratio on Public Benchmark.}
Table \ref{tab:compression_ratio} presents the CR of all evaluated compressors. DeLog achieves the highest CR on 15 out of 16 datasets, establishing a new state-of-the-art. On average, DeLog outperforms the next-best competitor, Denum, by 9.3\% (88.056 vs. 80.568). The performance gains are particularly notable on datasets like Zookeeper (+14.5\%), HealthApp (+25.8\%), and Windows (+12.5\%), where DeLog's superior pattern-based grouping proves more effective than Denum's numeric-focused approach. The only exception is the Android dataset, where Denum’s strategy of globally grouping numbers provides a slight advantage. DeLog-L, a variant designed for decompression speed, still achieves a competitive average CR of 69.991, outperforming LogReducer.

\begin{table}[ht]
\centering
\caption{The Compression Speed (MB/s) of Baselines. The best result is in \textbf{bold}.}
\label{tab:compression_speed}
\resizebox{\columnwidth}{!}{
\begin{tabular}{lccccc}
\toprule
Dataset & LogReducer & LogShrink & Denum & DeLog & DeLog-L \\
\hline
Android     & 3.096  & 0.448  & 15.427 & \textbf{25.232} & 22.423  \\
Apache      & 0.665   & 0.292  & 3.881  & \textbf{7.842}  & 6.325  \\
BGL         & 12.020  & 1.623  & 16.189 & \textbf{27.890} & 22.341  \\
Hadoop      & 2.976  & 0.508  & 17.449 & \textbf{32.393} & 29.784  \\
HDFS        & 14.280  & 1.812  & 17.444 & \textbf{31.232} & 29.932  \\
HealthApp   & 0.681   & 1.220  & 10.550 & \textbf{19.523} & 14.032  \\
HPC         & 3.850   & 1.430  & 10.617 & \textbf{20.081} & 16.849  \\
Linux       & 0.159   & 0.077  & 2.394  & \textbf{3.893}  & 3.423  \\
Mac         & 0.620   & 0.210  & 5.018  & \textbf{10.423} & 10.042  \\
OpenSSH     & 6.656  & 1.365  & 14.919 & \textbf{30.683} & 28.823  \\
OpenStack   & 5.516  & 1.142  & 9.735 & \textbf{14.978} & 12.734  \\
Proxifier   & 0.356   & 0.060  & 2.679  & 4.231  & \textbf{4.534}  \\
Spark       & 12.292  & 1.320  & 16.828 & \textbf{28.342} & 26.442 \\
Thunderbird & 8.435  & 1.321  & 13.460 & \textbf{29.181} & 28.712  \\
Windows     & 13.138  & 1.931  & 20.617 & 46.238 & \textbf{46.374} \\
Zookeeper   & 1.331   & 0.432  & 5.466  & \textbf{10.003} & 8.321  \\ 
\hline
Average     & 5.379  & 0.950  & 11.417 & \textbf{21.394} & 19.435  \\ 
\bottomrule
\end{tabular}
}
\end{table}

\textbf{Compression Speed on Public Benchmark.}
As shown in Table \ref{tab:compression_speed}, DeLog and DeLog-L dramatically outperform all baselines in compression speed. DeLog achieves an average speed of 21.4 MB/s, which is 87\% faster than the next-best competitor, Denum (11.4 MB/s), and over 22 times faster than LogShrink. DeLog-L maintains a commanding lead with an average speed of 19.4 MB/s. This significant speed advantage is a direct result of our lightweight, single-pass pattern-grouping design, which avoids the heavy computational overhead of traditional parsing or iterative analysis used by other tools.

\begin{table}[ht]
\centering
\caption{The Decompression Speed (MB/s) of Baselines on Public Benchmark. The best result is in \textbf{bold}.}
\label{tab:decompression_speed}
\resizebox{\columnwidth}{!}{
\begin{tabular}{lccccc}
\toprule
Dataset & LogReducer & LogShrink & Denum & DeLog & DeLog-L \\
\hline
Android     & 1.084  & 0.162  & 135.982 & 123.230 & \textbf{221.034}  \\
Apache      & 1.516   & 0.532  & 24.724  & 36.434  & \textbf{59.342}  \\
BGL         & 7.482  & 0.872  & 196.472 & 134.193 & \textbf{276.694}  \\
Hadoop      & 3.055   & 0.189  & 66.891 & 103.101 & \textbf{223.342} \\
HDFS        & 8.674  & 0.881  & 172.33 & 181.481 & \textbf{294.743}  \\
HealthApp   & 2.071   & Error  & 28.601 & 69.543 & \textbf{134.534}  \\
HPC         & 2.170   & 1.107  & 76.91 & 78.203 & \textbf{140.324}  \\
Linux       & 0.233   & 0.096  & 7.77  & 22.004  & \textbf{29.231}  \\
Mac         & 0.493   & 0.123  & 26.991  & 45.943 & \textbf{92.439}  \\
OpenSSH     & 4.404  & 0.730  & 173.24 & 143.231 & \textbf{243.234}  \\
OpenStack   & 6.546  & 0.866  & 67.512 & 89.782 & \textbf{153.423}  \\
Proxifier   & 0.083   & 0.071  & 7.879  & 20.799  & \textbf{43.433} \\
Spark       & 3.781  & 0.520  & 197.253 & 185.762 & \textbf{275.839}  \\
Thunderbird & 3.556  & 0.437  & 112.234 & 156.234 & \textbf{327.481}   \\
Windows     & 11.487  & 0.823  & 233.171 & 282.328 & \textbf{548.323}  \\
Zookeeper   & 0.456   & 0.733  & 30.532  & 38.932 & \textbf{78.343}  \\ 
\hline
Average     & 3.503  & 0.476  & 97.406 & 98.200 & \textbf{177.605}  \\ 
\bottomrule
\end{tabular}
}
\end{table}


\textbf{Decompression Speed on Public Benchmark.}
The decompression performance, detailed in Table \ref{tab:decompression_speed}, further underscores the efficiency of our design. DeLog-L sets a new state-of-the-art, reaching an average decompression speed of 177.6 MB/s, which is 82\% faster than Denum (97.4 MB/s). The standard DeLog is also highly competitive, performing on par with Denum at 98.2 MB/s. In contrast, other parser-based tools like LogReducer and LogShrink exhibit highly asymmetric performance, with decompression being their bottleneck. The rapid decompression of DeLog is enabled by its structured pattern signatures, which allow for a simple and direct reconstruction process without complex reverse logic. Notably, LogShrink's failure to decompress the HealthApp dataset also highlights the superior robustness of our approach.

\begin{tcolorbox}[colback=gray!10, colframe=black, sharp corners, boxrule=0.8pt]
\textbf{Summary for Q3:} DeLog achieves state-of-the-art compression ratios, outperforming competitors on 15 of 16 datasets. Both DeLog and its lightweight variant, DeLog-L, deliver significantly faster compression and decompression speeds than all baselines, demonstrating a superior balance of compression effectiveness and performance.
\end{tcolorbox}

\subsection{Q4: The Performance of DeLog on Production Logs}

To validate its real-world effectiveness, we evaluated DeLog against the same baselines on 10 large-scale production log datasets from ByteDance.

\textbf{Compression Ratio on Production Logs.}
The results, presented in Table \ref{tab:prod_cr}, underscore the limitations of existing tools and the significant advantage of DeLog. DeLog achieves the highest CR on all 10 production datasets, demonstrating its superior adaptability to complex, real-world data. On average, DeLog's CR of 26.3 is 38\% higher than the next-best competitor, Denum (19.1). Critically, both LogReducer and LogShrink failed to process 30\% of the datasets, crashing due to the logs' complexity. This highlights not only DeLog's superior compression but also its essential robustness for production environments.

\begin{figure*}[t]
	\centering
		\includegraphics[width=2\columnwidth]{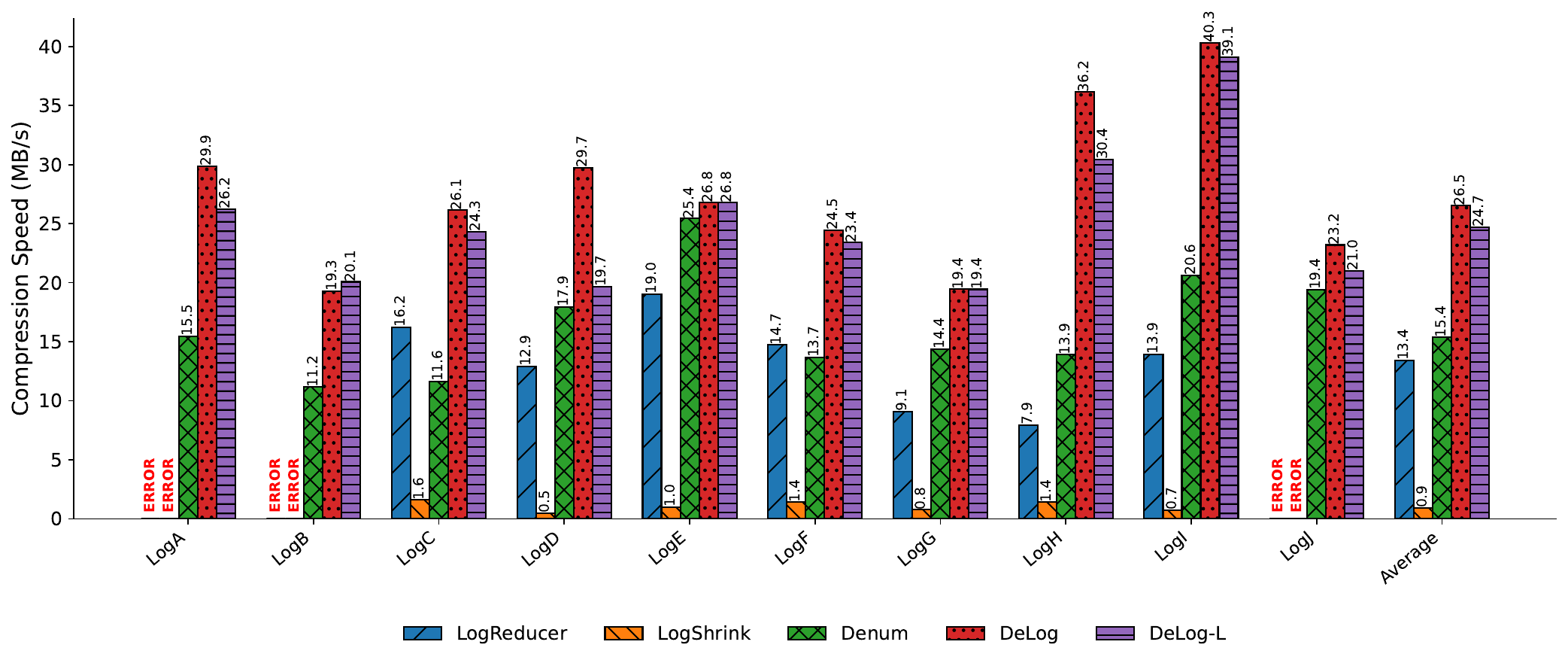}
	  \caption{Compression speed of baselines on Production Logs.}\label{com_speed_production}
\end{figure*}

\textbf{Compression Speed on Production Logs.}
As shown in Figure \ref{com_speed_production}, DeLog's performance advantage extends to speed. It is the fastest compressor on 9 out of 10 datasets. With an average speed of 26.5 MB/s, DeLog is 73\% faster than Denum (15.3 MB/s) and orders of magnitude faster than the parser-based methods. This validates that our lightweight, single-pass design avoids the performance bottlenecks that plague other tools when faced with the variable density and length of production logs. 

\textbf{Decompression Speed on Production Logs.}
The decompression results in Table \ref{tab:prod_ds} confirm the overall efficiency of our approach. DeLog-L delivers the fastest average decompression speed at 201.5~MB/s, outperforming Denum by 19\%. The standard DeLog is also highly competitive. This contrasts sharply with the extreme performance asymmetry of tools like LogReducer and LogShrink. Their slow decompression speeds translate to an unacceptable user experience; for example, waiting over half an hour just to access a 1~GB log file is impractical for any interactive use case. DeLog's simple and direct reconstruction mechanism, enabled by its structured pattern signatures, ensures rapid data access, providing both speed and robustness across all datasets.

\begin{tcolorbox}[colback=gray!10, colframe=black, sharp corners, boxrule=0.8pt]
\textbf{Summary for Q4:} On real-world production logs, DeLog proves both robust and efficient. It processes all datasets without failure where competitors crash, while also delivering state-of-the-art compression ratios and significantly faster compression and decompression speeds.
\end{tcolorbox}

\begin{table}[ht]
\centering
\caption{The Compression Ratio (CR) on Production Logs. The best result is in \textbf{bold}.}
\label{tab:prod_cr}
\resizebox{\columnwidth}{!}{
\begin{tabular}{lccccc}
\toprule
Dataset & LogReducer & LogShrink & Denum & DeLog & DeLog-L \\
\midrule
LogA & \textcolor{red}{\texttt{ERROR}} & \textcolor{red}{\texttt{ERROR}} & 14.629 & \textbf{19.487} & 19.239 \\
LogB & \textcolor{red}{\texttt{ERROR}} & \textcolor{red}{\texttt{ERROR}} & 13.465 & \textbf{21.834} & 21.244 \\
LogC & 21.108 & 20.900 & 19.014 & \textbf{28.611} & 22.086 \\
LogD & 11.235 & 9.932 & 13.897 & 12.510 & \textbf{12.464} \\
LogE & 10.987 & 11.020 & 11.051 & \textbf{11.144} & 11.130 \\
LogF & 26.390 & 27.940 & 28.091 & \textbf{47.444} & 46.616 \\
LogG & 33.953 & 30.031 & 22.022 & 40.273 & \textbf{40.621} \\
LogH & 27.942 & 25.982 & 27.321 & \textbf{38.311} & 36.961 \\
LogI & 32.030 & 35.902 & 29.299 & \textbf{54.312} & 46.329 \\
LogJ & \textcolor{red}{\texttt{ERROR}} & \textcolor{red}{\texttt{ERROR}} & 21.782 & \textbf{29.042} & 30.104 \\
\midrule
Average & 23.378 & 23.101 & 19.057 & \textbf{26.297} & 25.680 \\
\bottomrule
\end{tabular}}
\end{table}


\begin{table}[ht]
\centering
\caption{The Decompression Speed (MB/s) on Production Logs. The best result is in \textbf{bold}.}
\label{tab:prod_ds}
\resizebox{\columnwidth}{!}{
\begin{tabular}{lccccc}
\toprule
Dataset & LogReducer & LogShrink & Denum & DeLog & DeLog-L \\
\hline
LogA & \textcolor{red}{\texttt{ERROR}} & \textcolor{red}{\texttt{ERROR}} & 169.701 & 107.944 & \textbf{198.652} \\
LogB & \textcolor{red}{\texttt{ERROR}} & \textcolor{red}{\texttt{ERROR}} & 162.965 & 117.054 & \textbf{178.740} \\
LogC & 8.923 & 0.540 & 185.570 & 96.902 & \textbf{195.332} \\
LogD & 4.543 & 0.432 & 156.598 & 150.900 & \textbf{170.002} \\
LogE & 7.543 & 0.194 & 176.402 & 128.877 & \textbf{210.411} \\
LogF & 7.040 & 1.034 & 205.190 & 144.094 & \textbf{195.932} \\
LogG & 6.090 & 0.780 & 130.303 & 123.790 & \textbf{195.320} \\
LogH & 8.903 & 1.390 & 146.492 & 107.004 & \textbf{207.821} \\
LogI & 6.829 & 0.451 & 169.240 & 110.043 & \textbf{189.040} \\
LogJ & \textcolor{red}{\texttt{ERROR}} & \textcolor{red}{\texttt{ERROR}} & 190.432 & 166.069 & \textbf{273.902} \\
\hline
Average & 7.124 & 0.689 & 169.290 & 125.268 & \textbf{201.515} \\
\bottomrule
\end{tabular}
}
\end{table}

\subsection{Q5: Ablation Study on Feature Design}

\begin{figure}[H]
    \centering
    \includegraphics[width=\columnwidth]{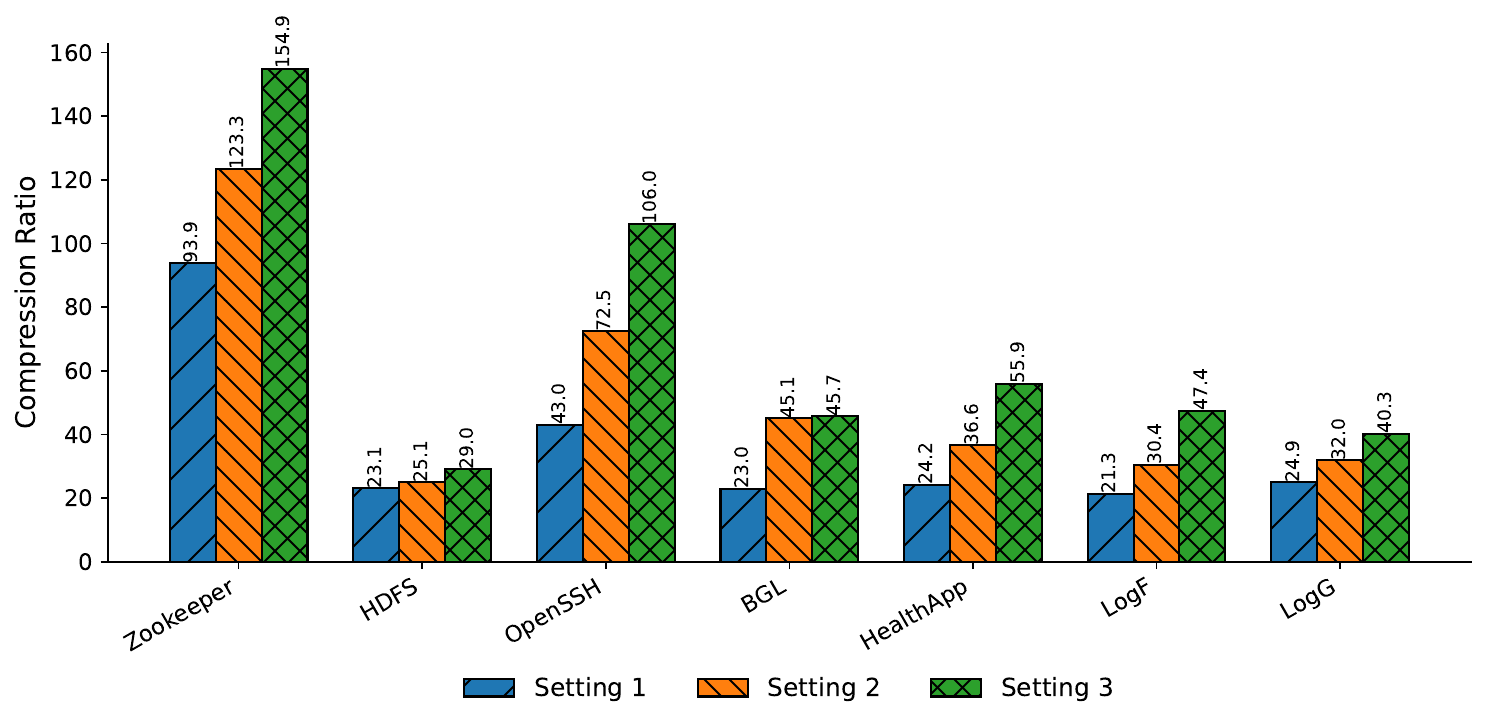}
    \caption{Ablation Study on the Impact of Different Feature Sets on Compression Ratio. We compare three settings: Setting 1 (Binary Grouping), a baseline that only distinguishes keywords from variables; Setting 2 (Intrinsic Structure), which adds analysis of the token's intrinsic structure; and Setting 3 (Intrinsic + External), the full DeLog model.}
    \label{fig:ablation_study}
\end{figure}

To validate the effectiveness of our multi-faceted feature design, we conducted an ablation study isolating the impact of intrinsic structure and external context on 7 randomly selected datasets. Figure~\ref{fig:ablation_study} shows the results across three settings.

We start with a baseline, Setting 1, which mimics a simple parser by performing only a binary grouping of keywords and variables. This yields the lowest CR across all datasets. By incorporating the token's intrinsic structure (Setting 2), we observe a significant performance improvement. For instance, on Zookeeper, the CR jumps from 93.86 to 123.34, and on HealthApp, from 24.19 to 36.63. This demonstrates that analyzing a token's intrinsic structure is critical for creating more effective, homogeneous groups.
Finally, the full DeLog model, Setting 3, which synthesizes both intrinsic structure and external context, achieves the best performance. The addition of external context provides another substantial boost. This highlights how contextual keywords are vital for disambiguating otherwise similar tokens. 

\begin{tcolorbox}[colback=gray!10, colframe=black, sharp corners, boxrule=0.8pt]
\textbf{Summary for Q5:} Each module of DeLog contributes to its compression ratio.
\end{tcolorbox}

\subsection{Scalability and CPU Efficiency}

Beyond raw compression performance, a practical log compressor must demonstrate high scalability and resource efficiency and to be viable in production. Our evaluation compares DeLog against a general-purpose compressor, LZMA, and a leading log-specific compressor, Denum. The tests span three large (>1m) public datasets (HDFS, BGL, Android) and four multi-gigabyte in-house production datasets (LogC-F), with thread counts varying from 1 to 16 to assess parallel performance.

\begin{figure}[t]
  \centering
  \includegraphics[width=\columnwidth]{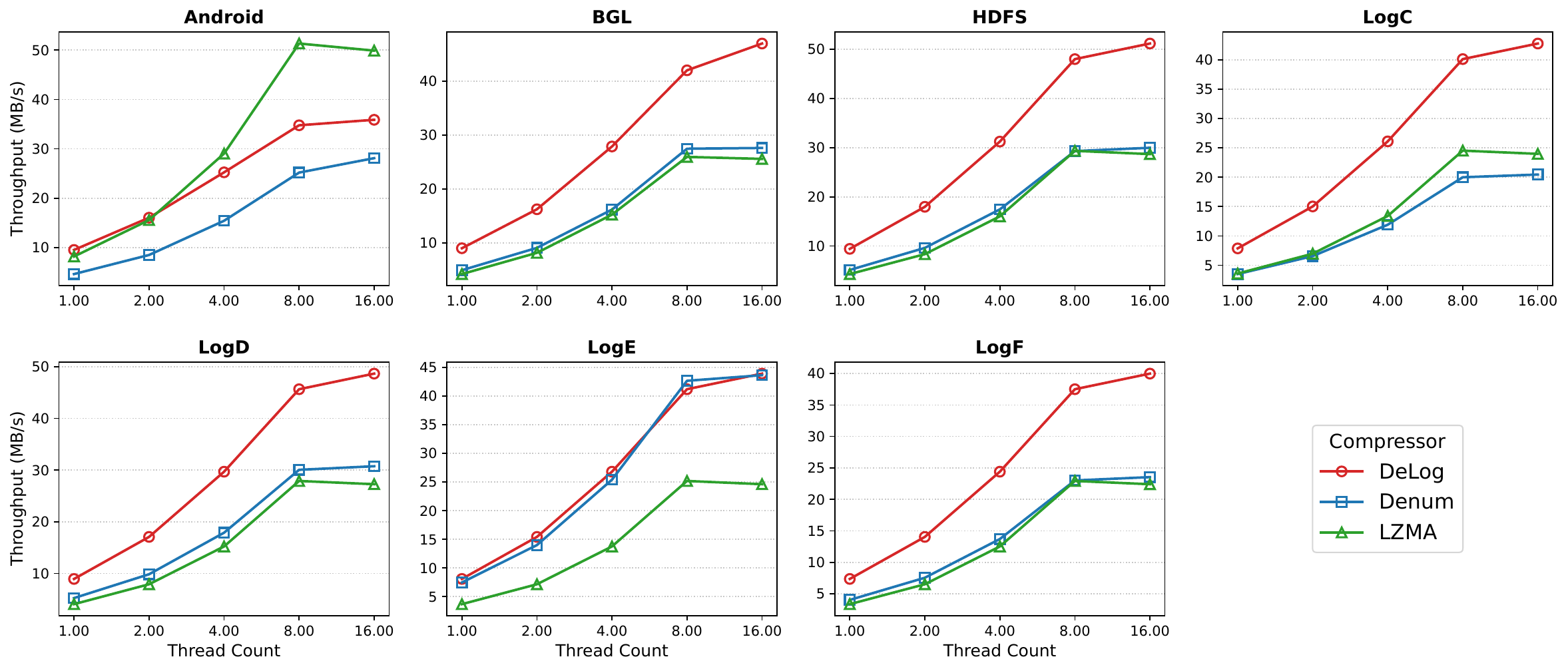}
  \caption{Scalability of compression throughput. DeLog consistently maintains the highest throughput across all thread counts due to its efficient core algorithm, while all compressors show similar scaling trends from leveraging data parallelism.}
  \label{fig:scalability}
\end{figure}

\textbf{Scalability}. Figure~\ref{fig:scalability} illustrates the compression throughput as a function of thread count. All evaluated compressors leverage data parallelism, resulting in similar scaling trends. The key differentiator, however, is not the scaling trend but the \textbf{absolute throughput}. 
DeLog consistently outperforms the baselines at every level of parallelism. For instance, on the BGL dataset, its throughput scales effectively from 10.2~MB/s (1~thread) to 44.7~MB/s (16~threads), always maintaining a significant lead. This sustained performance advantage stems directly from its superior singe-pass algorithmic efficiency. 

\begin{figure}[t]
  \centering
  \includegraphics[width=\linewidth]{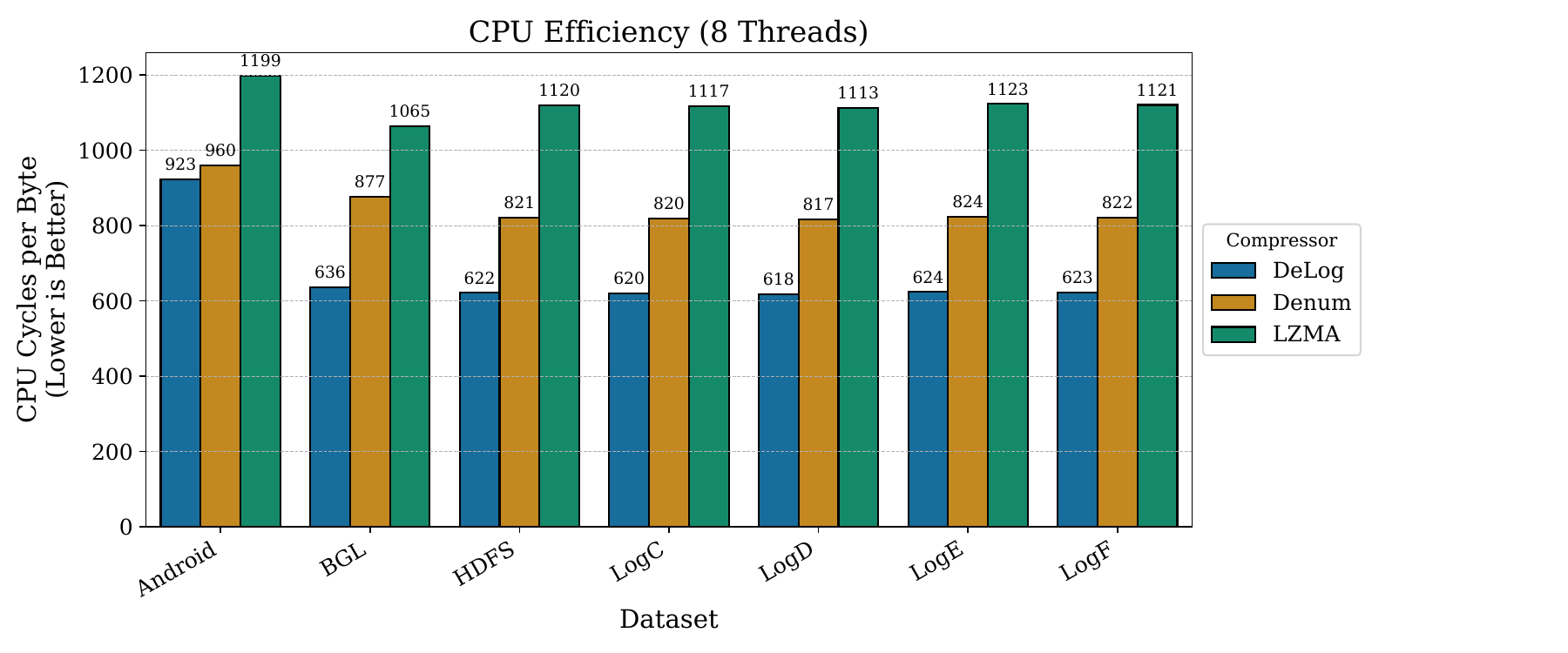}
  \caption{
    CPU efficiency comparison at 8 threads across all datasets. 
    The Y-axis shows the number of CPU cycles required to process a single byte of log data. 
    \textbf{Lower values indicate higher algorithmic efficiency.}
  }
  \label{fig:cpu-efficiency}
\end{figure}

\textbf{CPU Efficiency}. We first examine peak memory usage. On average across all datasets (8 threads), DeLog consumes 1172 MB of peak memory, whereas Denum requires 1320 MB. As expected for log-specific compressors that maintain structural information, these memory footprints are higher than that of general-purpose compressors like LZMA (avg. 661 MB). A truly efficient system must also perform its computational work economically. To quantify this, we analyze the CPU cycles per byte, a metric reflecting the intrinsic computational cost of an algorithm where lower values indicate higher efficiency. Figure \ref{fig:cpu-efficiency} presents this analysis for each compressor. The results unequivocally highlight DeLog's superior algorithmic efficiency. Across all seven datasets, DeLog consistently requires fewer CPU cycles per byte than Denum. On the BGL dataset, for example, DeLog consumes only 636 cycles/byte, making it 27.5\% more computationally efficient than Denum (877 cycles/byte). This proves that DeLog's high throughput stems from a fundamentally more efficient core algorithm, not just aggressive parallelization. Therefore, DeLog strikes an optimal balance, achieving not only higher throughput but also superior computational and memory efficiency compared to LZMA and Denum, making it a robust and well-engineered solution for large-scale deployments.

\section{Related Work}


\textbf{Log Compressors.}  
In addition to parer-based log compression, there are log compressors that apply unified compression strategies
to both templates and variables, such as the LogAchieve \cite{christensen2013adaptive}, Cowic \cite{lin2015cowic}, and MLC \cite{feng2016mlc}. Cowic prioritizes rapid data access over maximizing the compression ratio. MLC directly targets cross-entry redundancy through block-level deduplication. LogArchive uses similarity function
and sliding windows to divide log entries into different buckets and compresses buckets together to improve compression
ratio. Unlike LogArchive's coarse-grained, entry-level bucketing, our approach achieves fine-grained, pattern-based grouping at the token level, enabling more precise and effective data stream separation without log parsing. Specialized compressors have also been developed for structured formats. For example, uSlope \cite{wang2024muslope} targets JSON logs, where its notion of "semi-structured" data refers to varying schemas, a concept not applicable to the template/variable dichotomy in plain-text logs. Orthogonally, another class of compressors, including CLP \cite{rodrigues2021clp}, Loggrep \cite{wei2023loggrep}, and uSlope, prioritize search efficiency over storage savings. This focus often results in compression ratios lower than even general-purpose tools like LZMA. While search is not our primary goal, supplementary experiments show that DeLog's compression ratio is more than 2x higher than these search-optimized systems.

\textbf{Log Parsers.} Log parsers can be categorized into heuristic-based \cite{he2017drain,fu2022investigating,wang2022spine}, clustering-based \cite{fu2009execution, shima2016length}, frequent pattern mining-based \cite{dai2020logram,yu2023brain}, neural network-based \cite{huo2023semparser,nedelkoski2021self,liu2022uniparser, yu2023log}, and large language model-based \cite{xu2023prompting,jiang2023llmparser, le2023evaluation, le2023log}.
Heuristic-based methods extract log templates by leveraging rules observed within logs, while frequent pattern mining techniques focus on extracting recurring patterns, such as words or n-grams, as potential template candidates. Clustering methods group logs based on specific features to derive templates. Neural network-based methods apply various neural network architectures, while large language model-based methods leverage prompt engineering to guide large language models.

\section{Future Work}

While our syntactic approach is effective, the next frontier is to incorporate semantic understanding into our pattern-based grouping. Currently, our system cannot recognize that keywords like \texttt{taskID} and \texttt{TID} are semantically equivalent, thus missing opportunities for cross-keyword compression. Large Language Models (LLMs) offer a powerful path to bridge this semantic gap. However, the primary research challenge lies in mitigating the prohibitive latency and computational overhead of LLMs. Future work could explore hybrid models that combine our fast, syntactic grouper with a selective, on-demand semantic engine—perhaps aided by a query cache—to handle new or ambiguous patterns. This would evolve log compression from purely structural matching to a more intelligent,

\section{Conclusion}

This paper provides a new perspective on log compression by revealing a fundamental misalignment between the goals of traditional log parsing and the requirements of efficient compression. Our empirical study shows that pursuing higher parsing accuracy does not guarantee better compression. Instead, we establish that the key to log compression lies in pattern-based grouping, a principle focused on creating low-entropy and highly compressible groups.
Guided by this principle, we designed DeLog, a novel framework that forgoes traditional parsing to directly implement pattern-based grouping.



\section*{Availability}

All source code related to this paper is available at~\cite{delog}.

\bibliographystyle{plain}
\bibliography{main.bib}

@String{Computing = "Computing" }

@String{Computer = "{IEEE} Computer" }

@String{Springer = "Springer-Verlag" }

@article{he2020loghub,
  title={Loghub: A large collection of system log datasets towards automated log analytics},
  author={He, Shilin and Zhu, Jieming and He, Pinjia and Lyu, Michael R},
  journal={arXiv preprint arXiv:2008.06448},
  year={2020}
}

@article{he2021survey,
  title={A survey on automated log analysis for reliability engineering},
  author={He, Shilin and He, Pinjia and Chen, Zhuangbin and Yang, Tianyi and Su, Yuxin and Lyu, Michael R},
  journal={ACM computing surveys (CSUR)},
  volume={54},
  number={6},
  pages={1--37},
  year={2021},
  publisher={ACM New York, NY, USA}
}

@inproceedings{he2016experience,
  title={Experience report: System log analysis for anomaly detection},
  author={He, Shilin and Zhu, Jieming and He, Pinjia and Lyu, Michael R},
  booktitle={2016 IEEE 27th international symposium on software reliability engineering (ISSRE)},
  pages={207--218},
  year={2016},
  organization={IEEE}
}

@inproceedings{wei2021feasibility,
  title={On the Feasibility of Parser-based Log Compression in $\{$Large-Scale$\}$ Cloud Systems},
  author={Wei, Junyu and Zhang, Guangyan and Wang, Yang and Liu, Zhiwei and Zhu, Zhanyang and Chen, Junchao and Sun, Tingtao and Zhou, Qi},
  booktitle={19th USENIX Conference on File and Storage Technologies (FAST 21)},
  pages={249--262},
  year={2021}
}

@misc{lzmalink,
    title={lzma source code repositry},
    note={\url{https://7-zip.org/sdk.html}},
    year={2025-9-16}
}

@misc{gziplink,
    title={gzip source code repositry},
    note={\url{https://git.savannah.gnu.org/cgit/gzip.git}},
    year={2025-9-16}
}

@misc{delog,
    title={Delog code repositry},
    note={\url{https://github.com/gaiusyu/Delog}},
    year={2025-9-16}
}

@misc{brotil,
    title={brotil code repositry},
    note={\url{https://github.com/google/brotli}},
    year={2025-9-16}
}

@inproceedings{li2024logshrink,
  title={Logshrink: Effective log compression by leveraging commonality and variability of log data},
  author={Li, Xiaoyun and Zhang, Hongyu and Le, Van-Hoang and Chen, Pengfei},
  booktitle={Proceedings of the 46th IEEE/ACM International Conference on Software Engineering},
  pages={1--12},
  year={2024}
}

@inproceedings{liu2019logzip,
  title={Logzip: Extracting hidden structures via iterative clustering for log compression},
  author={Liu, Jinyang and Zhu, Jieming and He, Shilin and He, Pinjia and Zheng, Zibin and Lyu, Michael R},
  booktitle={2019 34th IEEE/ACM International Conference on Automated Software Engineering (ASE)},
  pages={863--873},
  year={2019},
  organization={IEEE}
}

@inproceedings{rodrigues2021clp,
  title={$\{$CLP$\}$: Efficient and Scalable Search on Compressed Text Logs},
  author={Rodrigues, Kirk and Luo, Yu and Yuan, Ding},
  booktitle={15th $\{$USENIX$\}$ Symposium on Operating Systems Design and Implementation ($\{$OSDI$\}$ 21)},
  pages={183--198},
  year={2021}
}

@inproceedings{lin2015cowic,
  title={Cowic: A column-wise independent compression for log stream analysis},
  author={Lin, Hao and Zhou, Jingyu and Yao, Bin and Guo, Minyi and Li, Jie},
  booktitle={2015 15th IEEE/ACM International Symposium on Cluster, Cloud and Grid Computing},
  pages={21--30},
  year={2015},
  organization={IEEE}
}

@article{yao2021improving,
  title={Improving state-of-the-art compression techniques for log management tools},
  author={Yao, Kundi and Sayagh, Mohammed and Shang, Weiyi and Hassan, Ahmed E},
  journal={IEEE Transactions on Software Engineering},
  volume={48},
  number={8},
  pages={2748--2760},
  year={2021},
  publisher={IEEE}
}

@inproceedings{wei2023loggrep,
  title={LogGrep: Fast and Cheap Cloud Log Storage by Exploiting both Static and Runtime Patterns},
  author={Wei, Junyu and Zhang, Guangyan and Chen, Junchao and Wang, Yang and Zheng, Weimin and Sun, Tingtao and Wu, Jiesheng and Jiang, Jiangwei},
  booktitle={Proceedings of the Eighteenth European Conference on Computer Systems},
  pages={452--468},
  year={2023}
}

@inproceedings{feng2016mlc,
  title={MLC: An efficient multi-level log compression method for cloud backup systems},
  author={Feng, Bo and Wu, Chentao and Li, Jie},
  booktitle={2016 IEEE Trustcom/BigDataSE/ISPA},
  pages={1358--1365},
  year={2016},
  organization={IEEE}
}

@inproceedings{he2017drain,
  title={Drain: An online log parsing approach with fixed depth tree},
  author={He, Pinjia and Zhu, Jieming and Zheng, Zibin and Lyu, Michael R},
  booktitle={2017 IEEE international conference on web services (ICWS)},
  pages={33--40},
  year={2017},
  organization={IEEE}
}

@inproceedings{christensen2013adaptive,
  title={Adaptive log compression for massive log data.},
  author={Christensen, Robert and Li, Feifei},
  booktitle={SIGMOD Conference},
  pages={1283--1284},
  year={2013}
}

@inproceedings{zhu2019tools,
  title={Tools and benchmarks for automated log parsing},
  author={Zhu, Jieming and He, Shilin and Liu, Jinyang and He, Pinjia and Xie, Qi and Zheng, Zibin and Lyu, Michael R},
  booktitle={2019 IEEE/ACM 41st International Conference on Software Engineering: Software Engineering in Practice (ICSE-SEIP)},
  pages={121--130},
  year={2019},
  organization={IEEE}
}

@inproceedings{jiang2008abstracting,
  title={Abstracting execution logs to execution events for enterprise applications (short paper)},
  author={Jiang, Zhen Ming and Hassan, Ahmed E and Flora, Parminder and Hamann, Gilbert},
  booktitle={2008 The Eighth International Conference on Quality Software},
  pages={181--186},
  year={2008},
  organization={IEEE},
  doi={10.1109/QSIC.2008.50}
}

@inproceedings{fu2009execution,
  title={Execution anomaly detection in distributed systems through unstructured log analysis},
  author={Fu, Qiang and Lou, Jian-Guang and Wang, Yi and Li, Jiang},
  booktitle={2009 ninth IEEE international conference on data mining},
  pages={149--158},
  year={2009},
  organization={IEEE}
}

@inproceedings{liu2022uniparser,
  title={Uniparser: A unified log parser for heterogeneous log data},
  author={Liu, Yudong and Zhang, Xu and He, Shilin and Zhang, Hongyu and Li, Liqun and Kang, Yu and Xu, Yong and Ma, Minghua and Lin, Qingwei and Dang, Yingnong and others},
  booktitle={Proceedings of the ACM Web Conference 2022},
  pages={1893--1901},
  year={2022}
}

@article{shima2016length,
  title={Length matters: Clustering system log messages using length of words},
  author={Shima, Keiichi},
  journal={arXiv preprint arXiv:1611.03213},
  year={2016}
}

@inproceedings{nedelkoski2021self,
  title={Self-supervised log parsing},
  author={Nedelkoski, Sasho and Bogatinovski, Jasmin and Acker, Alexander and Cardoso, Jorge and Kao, Odej},
  booktitle={Machine Learning and Knowledge Discovery in Databases: Applied Data Science Track: European Conference, ECML PKDD 2020, Ghent, Belgium, September 14--18, 2020, Proceedings, Part IV},
  pages={122--138},
  year={2021},
  organization={Springer}
}

@article{yu2023brain,
  title={Brain: Log Parsing with Bidirectional Parallel Tree},
  author={Yu, Siyu and He, Pinjia and Chen, Ningjiang and Wu, Yifan},
  journal={IEEE Transactions on Services Computing},
  year={2023},
  publisher={IEEE}
}

@article{dai2020logram,
  title={Logram: Efficient Log Parsing Using $ n $ n-Gram Dictionaries},
  author={Dai, Hetong and Li, Heng and Chen, Che-Shao and Shang, Weiyi and Chen, Tse-Hsun},
  journal={IEEE Transactions on Software Engineering},
  volume={48},
  number={3},
  pages={879--892},
  year={2020},
  publisher={IEEE}
}

@article{le2023log,
  title={Log Parsing with Prompt-based Few-shot Learning},
  author={Le, Van-Hoang and Zhang, Hongyu},
  journal={arXiv preprint arXiv:2302.07435},
  year={2023}
}

@article{le2023evaluation,
  title={An Evaluation of Log Parsing with ChatGPT},
  author={Le, Van-Hoang and Zhang, Hongyu},
  journal={arXiv preprint arXiv:2306.01590},
  year={2023}
}

@article{jiang2023llmparser,
  title={Llmparser: A llm-based log parsing framework},
  author={Jiang, Zhihan and Liu, Jinyang and Chen, Zhuangbin and Li, Yichen and Huang, Junjie and Huo, Yintong and He, Pinjia and Gu, Jiazhen and Lyu, Michael R},
  journal={arXiv preprint arXiv:2310.01796},
  year={2023}
}

@inproceedings{fu2022investigating,
  title={Investigating and improving log parsing in practice},
  author={Fu, Ying and Yan, Meng and Xu, Jian and Li, Jianguo and Liu, Zhongxin and Zhang, Xiaohong and Yang, Dan},
  booktitle={Proceedings of the 30th ACM Joint European Software Engineering Conference and Symposium on the Foundations of Software Engineering},
  pages={1566--1577},
  year={2022}
}

@article{xu2023prompting,
  title={Prompting for Automatic Log Template Extraction},
  author={Xu, Junjielong and Yang, Ruichun and Huo, Yintong and Zhang, Chengyu and He, Pinjia},
  journal={arXiv preprint arXiv:2307.09950},
  year={2023}
}

@inproceedings{wang2022spine,
  title={SPINE: a scalable log parser with feedback guidance},
  author={Wang, Xuheng and Zhang, Xu and Li, Liqun and He, Shilin and Zhang, Hongyu and Liu, Yudong and Zheng, Lingling and Kang, Yu and Lin, Qingwei and Dang, Yingnong and others},
  booktitle={Proceedings of the 30th ACM Joint European Software Engineering Conference and Symposium on the Foundations of Software Engineering},
  pages={1198--1208},
  year={2022}
}

@inproceedings{lu2017log,
  title={Log-based abnormal task detection and root cause analysis for spark},
  author={Lu, Siyang and Rao, BingBing and Wei, Xiang and Tak, Byungchul and Wang, Long and Wang, Liqiang},
  booktitle={2017 IEEE International Conference on Web Services (ICWS)},
  pages={389--396},
  year={2017},
  organization={IEEE}
}

@inproceedings{yu2023log,
  title={Log Parsing with Generalization Ability under New Log Types},
  author={Yu, Siyu and Wu, Yifan and Li, Zhijing and He, Pinjia and Chen, Ningjiang and Liu, Changjian},
  booktitle={Proceedings of the 31st ACM Joint European Software Engineering Conference and Symposium on the Foundations of Software Engineering},
  pages={425--437},
  year={2023}
}

@inproceedings{yang2018nanolog,
  title={$\{$NanoLog$\}$: A Nanosecond Scale Logging System},
  author={Yang, Stephen and Park, Seo Jin and Ousterhout, John},
  booktitle={2018 USENIX Annual Technical Conference (USENIX ATC 18)},
  pages={335--350},
  year={2018},
}

@inproceedings{wang2024muslope,
  title={$\{$$\mu$Slope$\}$: High Compression and Fast Search on $\{$Semi-Structured$\}$ Logs},
  author={Wang, Rui and Gibson, Devin and Rodrigues, Kirk and Luo, Yu and Zhang, Yun and Wang, Kaibo and Fu, Yupeng and Chen, Ting and Yuan, Ding},
  booktitle={18th USENIX Symposium on Operating Systems Design and Implementation (OSDI 24)},
  pages={529--544},
  year={2024},

}

@article{khan2024impact,
  title={Impact of log parsing on deep learning-based anomaly detection},
  author={Khan, Zanis Ali and Shin, Donghwan and Bianculli, Domenico and Briand, Lionel C},
  journal={Empirical Software Engineering},
  volume={29},
  number={6},
  pages={139},
  year={2024},
  publisher={Springer}
}

@article{fu2023empirical,
  title={An empirical study of the impact of log parsers on the performance of log-based anomaly detection},
  author={Fu, Ying and Yan, Meng and Xu, Zhou and Xia, Xin and Zhang, Xiaohong and Yang, Dan},
  journal={Empirical Software Engineering},
  volume={28},
  number={1},
  pages={6},
  year={2023},
  publisher={Springer}
}

@inproceedings{yu2024unlocking,
  title={Unlocking the Power of Numbers: Log Compression via Numeric Token Parsing},
  author={Yu, Siyu and Wu, Yifan and Li, Ying and He, Pinjia},
  booktitle={Proceedings of the 39th IEEE/ACM International Conference on Automated Software Engineering},
  pages={919--930},
  year={2024}
}

@inproceedings{li2023did,
  title={Did we miss something important? studying and exploring variable-aware log abstraction},
  author={Li, Zhenhao and Luo, Chuan and Chen, Tse-Hsun and Shang, Weiyi and He, Shilin and Lin, Qingwei and Zhang, Dongmei},
  booktitle={2023 IEEE/ACM 45th International Conference on Software Engineering (ICSE)},
  pages={830--842},
  year={2023},
  organization={IEEE}
}

@inproceedings{jiang2024large,
  title={A large-scale evaluation for log parsing techniques: How far are we?},
  author={Jiang, Zhihan and Liu, Jinyang and Huang, Junjie and Li, Yichen and Huo, Yintong and Gu, Jiazhen and Chen, Zhuangbin and Zhu, Jieming and Lyu, Michael R},
  booktitle={Proceedings of the 33rd ACM SIGSOFT International Symposium on Software Testing and Analysis},
  pages={223--234},
  year={2024}
}

@inproceedings{khan2022guidelines,
  title={Guidelines for assessing the accuracy of log message template identification techniques},
  author={Khan, Zanis Ali and Shin, Donghwan and Bianculli, Domenico and Briand, Lionel},
  booktitle={Proceedings of the 44th International Conference on Software Engineering},
  pages={1095--1106},
  year={2022}
}

@inproceedings{zhang2019robust,
  title={Robust log-based anomaly detection on unstable log data},
  author={Zhang, Xu and Xu, Yong and Lin, Qingwei and Qiao, Bo and Zhang, Hongyu and Dang, Yingnong and Xie, Chunyu and Yang, Xinsheng and Cheng, Qian and Li, Ze and others},
  booktitle={Proceedings of the 2019 27th ACM joint meeting on European software engineering conference and symposium on the foundations of software engineering},
  pages={807--817},
  year={2019}
}

@inproceedings{huo2023semparser,
  title={Semparser: A semantic parser for log analytics},
  author={Huo, Yintong and Su, Yuxin and Lee, Cheryl and Lyu, Michael R},
  booktitle={2023 IEEE/ACM 45th International Conference on Software Engineering (ICSE)},
  pages={881--893},
  year={2023},
  organization={IEEE}
}

@article{chen2025tracezip,
  title={Tracezip: Efficient Distributed Tracing via Trace Compression},
  author={Chen, Zhuangbin and Pu, Junsong and Zheng, Zibin},
  journal={arXiv preprint arXiv:2502.06318},
  year={2025}
}

@article{jiang2024lilac,
  title={Lilac: Log parsing using llms with adaptive parsing cache},
  author={Jiang, Zhihan and Liu, Jinyang and Chen, Zhuangbin and Li, Yichen and Huang, Junjie and Huo, Yintong and He, Pinjia and Gu, Jiazhen and Lyu, Michael R},
  journal={Proceedings of the ACM on Software Engineering},
  volume={1},
  number={FSE},
  pages={137--160},
  year={2024},
  publisher={ACM New York, NY, USA}
}

@inproceedings{makanju2009clustering,
  title={Clustering event logs using iterative partitioning},
  author={Makanju, Adetokunbo AO and Zincir-Heywood, A Nur and Milios, Evangelos E},
  booktitle={Proceedings of the 15th ACM SIGKDD international conference on Knowledge discovery and data mining},
  pages={1255--1264},
  year={2009}
}

@inproceedings{vaarandi2015logcluster,
  title={Logcluster-a data clustering and pattern mining algorithm for event logs},
  author={Vaarandi, Risto and Pihelgas, Mauno},
  booktitle={2015 11th International conference on network and service management (CNSM)},
  pages={1--7},
  year={2015},
  organization={IEEE}
}

@inproceedings{nagappan2010abstracting,
  title={Abstracting log lines to log event types for mining software system logs},
  author={Nagappan, Meiyappan and Vouk, Mladen A},
  booktitle={2010 7th IEEE Working Conference on Mining Software Repositories (MSR 2010)},
  pages={114--117},
  year={2010},
  organization={IEEE}
}

@inproceedings{xiao2024free,
  title={Free: Towards More Practical Log Parsing with Large Language Models},
  author={Xiao, Yi and Le, Van-Hoang and Zhang, Hongyu},
  booktitle={Proceedings of the 39th IEEE/ACM International Conference on Automated Software Engineering},
  pages={153--165},
  year={2024}
}

@inproceedings{zhang2025scalalog,
  title={ScalaLog: Scalable Log-Based Failure Diagnosis Using LLM},
  author={Zhang, Lingzhe and Jia, Tong and Jia, Mengxi and Wu, Yifan and Liu, Hongyi and Li, Ying},
  booktitle={ICASSP 2025-2025 IEEE International Conference on Acoustics, Speech and Signal Processing (ICASSP)},
  pages={1--5},
  year={2025},
  organization={IEEE}
}

@inproceedings{xu2023hue,
  title={Hue: A user-adaptive parser for hybrid logs},
  author={Xu, Junjielong and Fu, Qiuai and Zhu, Zhouruixing and Cheng, Yutong and Li, Zhijing and Ma, Yuchi and He, Pinjia},
  booktitle={Proceedings of the 31st ACM Joint European Software Engineering Conference and Symposium on the Foundations of Software Engineering},
  pages={413--424},
  year={2023}
}

@article{zhang2025logbase,
  title={LogBase: A Large-Scale Benchmark for Semantic Log Parsing},
  author={Zhang, Chenbo and Xu, Wenying and Liu, Jinbu and Zhang, Lu and Liu, Guiyang and Guan, Jihong and Zhou, Qi and Zhou, Shuigeng},
  journal={Proceedings of the ACM on Software Engineering},
  volume={2},
  number={ISSTA},
  pages={2091--2112},
  year={2025},
  publisher={ACM New York, NY, USA}
}

\appendix
\section*{Appendix}

This supplementary document provides additional details and extended experimental results to complement our main paper, "DeLog: A Pattern-based Grouper for Compressing Production Logs."

The content is organized into two main parts:

\begin{itemize}
    \item \textbf{Section A} presents the full details of our in-depth empirical study on the relationship between log parsing accuracy and log compression. This section elaborates on the questions, methodology, evaluation metrics, and the detailed findings that led to our core insight: that effective pattern-based grouping, not formal parsing accuracy, is the true driver of high compression ratios.

    \item \textbf{Section B} contains an extended experimental analysis, comparing the performance of our proposed method, DeLog, against standard general-purpose compressors (\texttt{gzip}, \texttt{bzip2}, and \texttt{lzma}) on the public benchmark datasets. This analysis provides a broader context for our results, quantitatively demonstrating the significant advantages of a domain-specific approach for log compression in terms of both compression ratio and the balance between speed and effectiveness.
\end{itemize}

This material is intended to provide a more comprehensive background for interested readers and to support the reproducibility of our work.

\section{An Empirical Study of the Impact of Log Parsing Accuracy on Log Compression}\label{empirical}

We try to answer the following two research questions to identify the most promising direction for advancing log compression:

\begin{itemize}
    \item[\textbf{Q1:}] \textit{What is the correlation between log parsing accuracy and the final compression ratio?} A weak or inconsistent correlation would suggest a fundamental misalignment between the goals of parsing research and the needs of compression.

    \item[\textbf{Q2:}] \textit{If not accurate parsed format, what structural properties of a processed log make it highly compressible?} Answering this is essential for designing a new generation of truly compression-oriented pre-processors.
\end{itemize}

\subsection{Methodology.}
To assess the correlation between parsing and compression, we first establish a baseline by comparing the CR of parsed logs against raw logs. Next, we systematically measure the Spearman correlation coefficient between six different log parsing accuracy metrics and the final CR across multiple datasets and parsers. This will quantitatively determine if "more accurate" means "more compressible."

If the results from Q1 show a weak correlation, a qualitative deep-dive is necessary to identify the true drivers of compression. Our methodology for this is a two-step analysis: 1. Deconstruct the Parsing Pipeline: First, we isolate the contributions of the two main parsing stages: log formatting (splitting into fields) and template extraction. By measuring the CR after each stage, we can determine which step provides the most significant compression gains. 2. Analyze High-Variance Cases: Second, we analyze specific cases from our experiments where different parsers achieve dramatically different CRs on the same dataset. For these cases, we manually inspect the intermediate parsed formats to pinpoint the structural properties that lead to superior compression. The ultimate goal of this two-pronged analysis is to distill our observations into a generalizable principle that defines a truly compression-friendly log format.

\subsection{The Evaluation Metrics of Log Parsing}\label{metric_intro}

The most common definition of log parsing is the transformation of raw, unstructured logs into structured log templates, which is also referred to as log abstraction in some literature \cite{li2023did}. The quality of log parsers is typically measured by a set of standard metrics that evaluate correctness at either the log level or the template level.

The primary log-level metrics are Group Accuracy and Parsing Accuracy.
Group Accuracy (GA) assesses the correctness of log message grouping. It is calculated as the ratio of correctly grouped logs to the total number of logs \cite{zhu2019tools}. A single log is considered correctly grouped if and only if all logs belonging to its ground-truth template are clustered into the same group by the parser.
Parsing Accuracy (PA), also known as Word-level Parsing Accuracy (WLA), provides a finer-grained evaluation. It is defined as the ratio of correctly parsed logs to the total number of logs \cite{liu2022uniparser}. For PA, a log is considered correct only if every word within it is accurately classified as either a constant (part of the template) or a variable (a parameter).

To mitigate the issue where a high volume of repetitive logs can skew log-level metrics, template-level metrics were introduced.
Template Accuracy (TA) evaluates the quality of the set of generated templates instead of individual logs. It typically comprises three specific metrics \cite{khan2022guidelines,jiang2024large}: Precision Template Accuracy (PTA), the ratio of correctly identified templates to the total number of templates generated by the parser; Recall Template Accuracy (RTA), the ratio of correctly identified templates to the total number of templates in the ground truth; and the F1-score of Template Accuracy (FTA), which is their harmonic mean. A template is considered correct only if its content and its corresponding log group perfectly match the ground truth.
The F1-score of Group Accuracy (FGA) is another template-level F1-score metric \cite{jiang2024large}. Similar to TA, it is the harmonic mean of a precision and a recall score calculated at the template level. Its precision is the fraction of generated templates that are correctly grouped, and its recall is the fraction of ground-truth templates that are correctly found and grouped.

\subsection{Experimental Design}
\textbf{Dataset.} Our study requires large, labeled datasets to robustly evaluate parsing accuracy. The standard Loghub benchmark is insufficient due to its small labeled subsets (2k logs per dataset). We therefore use 12 diverse datasets from Loghub 2.0, which provides over 15 million labeled entries spanning various system types, including Distributed (e.g., Hadoop), Supercomputer (e.g., BGL), Operating (e.g., Android) and Server (e.g., Apache) systems.

\begin{table}[h]
    \centering
    \caption{Statistics of Loghub-2.0.}
    \resizebox{\columnwidth}{!}{
    \begin{tabular}{@{}p{2.1cm}p{1.8cm}p{1.8cm}p{1.8cm}@{}}
        \toprule
        System & Dataset & Templates & Annotated Logs \\ \midrule
        \multirow{4}{*}{\parbox{2.1cm}{\centering \textbf{Distributed systems}}} & Hadoop & 236 & 179,993 \\
        & HDFS & 46 & 11,167,740 \\
        & OpenStack & 48 & 207,632 \\
        & Zookeeper & 89 & 74,273 \\ \midrule
        
        \multirow{3}{*}{\parbox{2.1cm}{\centering \textbf{Super-computer systems}}} & BGL & 320 & 4,631,261 \\
        & HPC & 74 & 429,987 \\ 
        &  &  &  \\ \midrule
        
        \multirow{2}{*}{\parbox{2.1cm}{\centering \textbf{Operating systems}}} & Linux & 338 & 23,921 \\
        & Mac & 626 & 100,314 \\ \midrule
        
        \multirow{2}{*}{\parbox{2.1cm}{\centering \textbf{Server application}}} & Apache & 29 & 51,977 \\
        & OpenSSH & 38 & 638,946 \\ \midrule
        
        \multirow{2}{*}{\parbox{2.1cm}{\centering \textbf{Standalone software}}} & HealthApp & 156 & 212,394 \\
        & Proxifier & 11 & 21,320 \\ 
        \bottomrule
    \end{tabular}
    }\label{loghub2}
\end{table}

\textbf{Metrics.} For log parsing accuracy, we selected all existing log accuracy metrics, including Group Accuracy (GA), F1-score of Group Accuracy (FGA), Parsing Accuracy (PA), Precision of Template Accuracy (PTA), Recall of Template Accuracy (RTA), and F1-score of Template Accuracy (FTA). These metrics have been detailed in Section \ref{metric_intro}. 

The Spearman's Rank Correlation Coefficient (denoted as \( \rho \) or \( r_s \)) assesses the strength and direction of association between two ranked variables. 

Calculation:
\[
\rho = 1 - \frac{6 \sum d_i^2}{n(n^2 - 1)}
\]
where \( d_i \) is the difference between ranks of each pair and \( n \) is the number of pairs. According to prior research examining the correlation between log parsing and its downstream tasks, a Spearman’s correlation coefficient with an absolute value exceeding 0.7 is considered indicative of a strong correlation \cite{fu2023empirical, khan2024impact}.

\textbf{Experimental Setup.} For practicality, diversity, and observability, we selected log parsers from LogPAI that exhibit differences in log parsing accuracy and relatively fast parsing speeds. Although LLM-based log parsers \cite{jiang2024lilac, xu2023prompting, xiao2024free} have achieved superior performance, their processing speed is inefficient for log compression. As an alternative, we include the Ground Truth (GT) as a perfect log parser in our comparative experiments, where all accuracy metrics are equal to 1. Selected log Parsers include heuristic-based parsers (Drain \cite{he2017drain}, AEL \cite{jiang2008abstracting}), iterative partitioning-based parsers (IPLoM \cite{makanju2009clustering}), frequent pattern mining-based parsers (LogCluster \cite{vaarandi2015logcluster}, LFA \cite{nagappan2010abstracting}), and clustering-based parsers (Brain \cite{yu2023brain}). For encoding techniques, we employ a combined encoding backend for all parsers rather than using existing log compressors. This is reasonable because existing parser-based log compressors generally are extensions of one another. Our backend applies a common set of techniques—dictionary, delta, and elastic encoding—followed by a final LZMA compression stage. This design allows us to attribute any differences in the final compression ratio directly to the performance of the parsing accuracy under study.

\textbf{Environment.} We performed all experiments using a Linux server, equipped with a single Intel(R) Xeon(R) Platinum 8369HB CPU @ 3.30GHz, featuring 16 cores and 32 threads due to hyper-threading. The server is provisioned with 188GB of RAM and operates on Red Hat with Linux kernel 3.10.0.

\begin{figure}[t]
	\centering
		\includegraphics[width=\columnwidth]{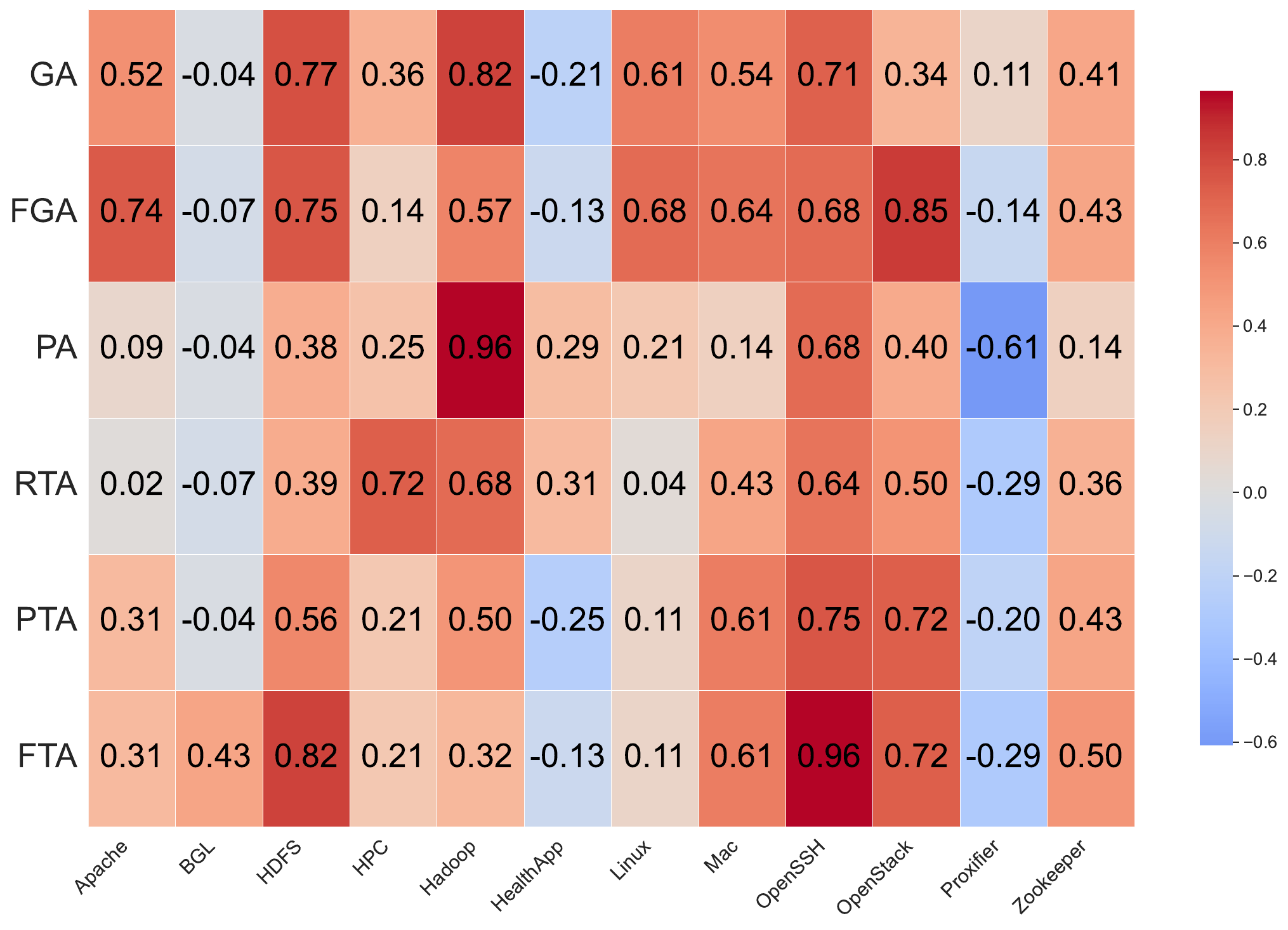}
	  \caption{Spearman Correlation Between Log Parsing Accuracy Metrics (i.e., GA, FGA, PA, RTA, PTA, FTA) and CR. The closer the color is to red, the stronger the positive correlation; conversely, the closer it is to blue, the stronger the negative correlation.}\label{spearman_fig}
\end{figure}

\begin{figure}[t]
	\centering
		\includegraphics[width=\columnwidth]{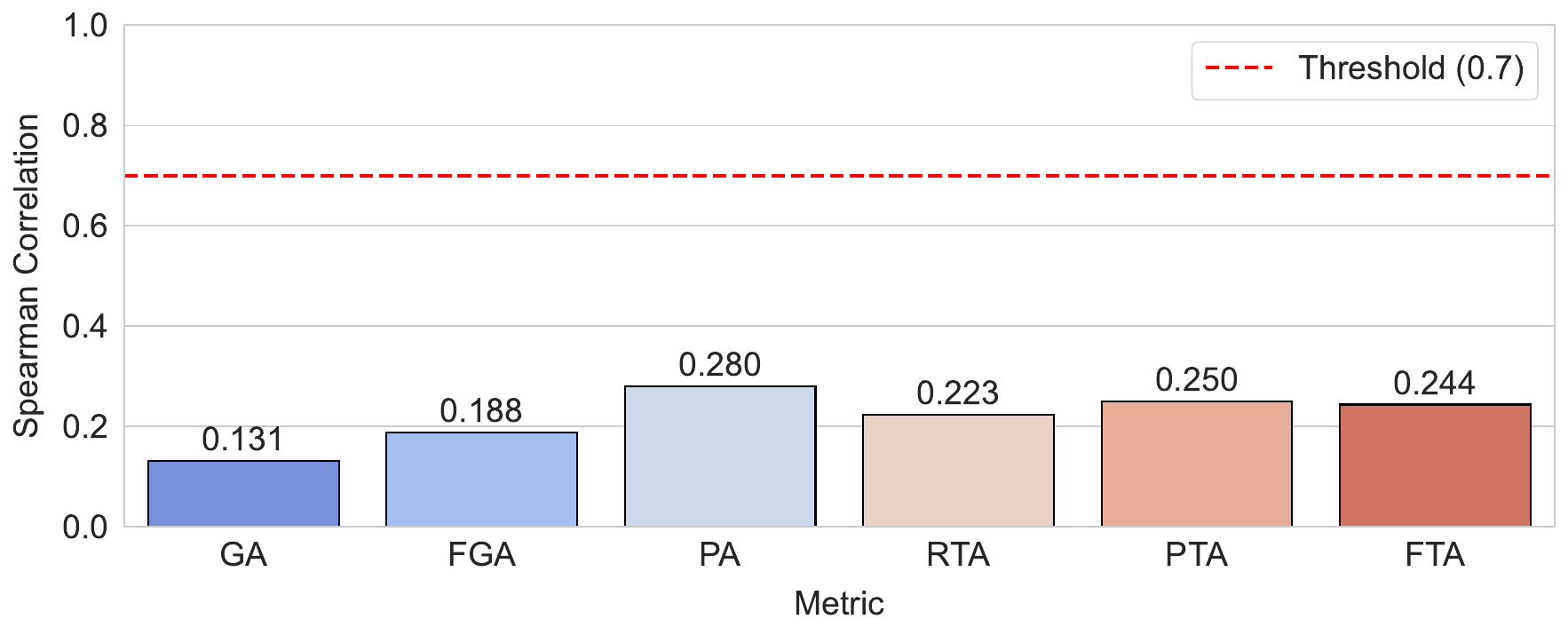}
	  \caption{Spearman Correlation Between Log Parsing Accuracy Metrics and CR (All Datasets).}\label{spearman_all}
\end{figure}


\begin{table*}[p] 
\centering 
\caption{Comparison of Log Parsing Accuracy Metrics and Compression Ratio Across Different Datasets and Parsers. The highest log parsing accuracy and CR in each dataset is highlighted with a gray background, and the lowest CR is underscored.} 
\label{tab:log_parser_compression_transposed} 
\begin{adjustbox}{width=\textwidth} 
\begin{tabular}{l|ccccccc|ccccccc} 
\toprule 
\multirow{2}{*}{Metric} & \multicolumn{7}{c}{Apache Raw:28.12 Formatted:37.11 } & \multicolumn{7}{c}{Proxifier Raw:20.77 Formatted:24.54} \\ 
\cmidrule(lr){2-8} \cmidrule(lr){9-15} 
& Drain & AEL & IPLoM & LFA & LogCluster & Brain & GT & Drain & AEL & IPLoM & LFA & LogCluster & Brain & GT \\ 
\midrule 
GA (\%) & \cellcolor{lightgray}100.0 & \cellcolor{lightgray} 100.0 & 99.3 & 80.5 & 54.7 & 99.7 & \cellcolor{lightgray} 100.0 & 69.2 & 97.4 & 98.3 & 35.2 & 66.1 & 52.1 & \cellcolor{lightgray} 100.0 \\ 
FGA (\%) & \cellcolor{lightgray} 100.0 & \cellcolor{lightgray} 100.0 & 90.0 & 87.3 & 00.3 & 93.3 & \cellcolor{lightgray} 100.0 & 20.6 & 66.7 & 75.0 & 40.0 & 00.1 & 73.7 & \cellcolor{lightgray} 100.0 \\ 
PA (\%) & 72.7 & 72.7 & 25.0 & 63.7 & 43.2 & 28.7 & \cellcolor{lightgray} 100.0 & 68.8 & 67.7 & 48.0 & 00.0 & 66.1 & 70.3 & \cellcolor{lightgray} 100.0 \\ 
RTA (\%) & 51.7 & 51.7 & 17.2 & 24.1 & 27.6 & 48.3 & \cellcolor{lightgray} 100.0 & 54.5 & 45.5 & 18.2 & 00.0 & 18.2 & 63.6 & \cellcolor{lightgray} 100.0 \\ 
PTA (\%) & 51.7 & 51.7 & 16.1 & 26.9 & 00.1 & 46.7 & \cellcolor{lightgray} 100.0 & 10.5 & 38.5 & 15.4 & 00.0 & 00.0 & 59.6 & \cellcolor{lightgray} 100.0 \\ 
FTA (\%) & 51.7 & 51.7 & 16.7 & 25.5 & 00.1 & 41.7 & \cellcolor{lightgray} 100.0 & 17.6 & 41.7 & 16.7 & 00.0 & 00.1 & 39.9 & \cellcolor{lightgray} 100.0 \\ 
CR & 49.70 & 49.99 & \cellcolor{lightgray} 50.09 & 49.16 & \underline{39.11} & 46.67 & 49.42 & 25.67 & 26.35 & \cellcolor{lightgray} 26.39 & 25.63 & 24.61 & 24.56 & \underline{24.55} \\ 
\midrule 
\multirow{2}{*}{Metric} & \multicolumn{7}{c}{Linux Raw:18.09 Formatted:20.93} & \multicolumn{7}{c}{Zookeeper Raw:27.55 Formatted:61.69} \\ 
\cmidrule(lr){2-8} \cmidrule(lr){9-15} 
& Drain & AEL & IPLoM & LFA & LogCluster & Brain & GT & Drain & AEL & IPLoM & LFA & LogCluster & Brain & GT \\ 
\midrule 
GA (\%) & 68.6 & 91.6 & 81.0 & 22.5 & 59.5 & 79.0 & \cellcolor{lightgray} 100.0 & 99.4 & 99.6 & 99.4 & 83.9 & 73.7 & 99.3 & \cellcolor{lightgray} 100.0 \\ 
FGA (\%) & 77.8 & 80.6 & 81.2 & 73.9 & 29.2 & 75.1 & \cellcolor{lightgray} 100.0 & 90.4 & 78.8 & 85.9 & 62.6 & 08.8 & 79.8 & \cellcolor{lightgray} 100.0 \\ 
PA (\%) & 11.1 & 08.2 & 05.5 & 03.3 & 02.3 & 09.8 & \cellcolor{lightgray} 100.0 & 84.3 & 84.2 & 79.1 & 34.6 & 72.8 & 82.3 & \cellcolor{lightgray} 100.0 \\ 
RTA (\%) & 29.9 & 21.9 & 11.2 & 00.8 & 18.3 & 31.7 & \cellcolor{lightgray} 100.0 & 57.3 & 51.7 & 40.4 & 20.2 & 38.2 & 62.9 & \cellcolor{lightgray} 100.0 \\ 
PTA (\%) & 22.9 & 21.5 & 11.4 & 07.8 & 03.9 & 31.7 & \cellcolor{lightgray} 100.0 & 66.2 & 42.2 & 48.6 & 24.3 & 03.3 & 59.6 & \cellcolor{lightgray} 100.0 \\ 
FTA (\%) & 25.9 & 21.7 & 11.3 & 07.9 & 06.4 & 28.0 & \cellcolor{lightgray} 100.0 & 61.4 & 46.5 & 44.2 & 22.1 & 06.0 & 61.2 & \cellcolor{lightgray} 100.0 \\ 
CR & 22.79 & 22.98 & \cellcolor{lightgray} 23.35 & \underline{20.11} & 20.95 & 20.35 & 21.81 & 77.05 & 76.81 & 75.86 & \underline{73.53} &74.60 & \cellcolor{lightgray}78.08 & 76.72 \\ 
\midrule 
\multirow{2}{*}{Metric} & \multicolumn{7}{c}{Hadoop Raw:21.76 Formatted:41.64} & \multicolumn{7}{c}{HealthApp Raw:17.73 Formatted:22.23} \\ 
\cmidrule(lr){2-8} \cmidrule(lr){9-15} 
&Drain & AEL & IPLoM & LFA & LogCluster & Brain & GT & Drain & AEL & IPLoM & LFA & LogCluster & Brain & GT \\ 
\midrule 
GA (\%) & 92.1 & 82.3 & 91.6 & 82.7 & 48.1 & 56.3 & \cellcolor{lightgray} 100.0 & 86.2 & 72.5 & 97.7 & 80.4 & 73.1 & 98.0 & \cellcolor{lightgray} 100.0 \\ 
FGA (\%) & 78.5 & 11.3 & 79.7 & 65.0 & 00.5 & 20.3 & \cellcolor{lightgray} 100.0 & 01.0 & 00.8 & 88.0 & 00.8 & 00.6 & 90.2 & \cellcolor{lightgray} 100.0 \\ 
PA (\%) & 54.1 & 53.5 & 16.2 & 43.2 & 30.7 & 28.8 & \cellcolor{lightgray} 100.0 & 31.2 & 31.1 & 16.3 & 30.6 & 40.5 & 14.8 & \cellcolor{lightgray} 100.0 \\ 
RTA (\%) & 45.3 & 42.8 & 26.7 & 13.1 & 14.0 & 28.8 & \cellcolor{lightgray} 100.0 & 35.3 & 31.4 & 25.6 & 25.6 & 25.0 & 61.2 & \cellcolor{lightgray} 100.0 \\ 
PTA (\%) & 33.2 & 03.1 & 22.0 & 10.7 & 00.1 & 41.7 & \cellcolor{lightgray} 100.0 & 00.2 & 00.2 & 25.0 & 00.1 & 00.1 & 39.9 & \cellcolor{lightgray} 100.0 \\ 
FTA (\%) & 38.4 & 05.8 & 24.1 & 11.8 & 00.2 & 39.9 & \cellcolor{lightgray} 100.0 & 00.4 & 00.3 & 25.3 & 00.3 & 00.2 & 39.9 & \cellcolor{lightgray} 100.0 \\ 
CR & 50.39 & 50.04 & 45.59 & 49.26 & 44.04 & \underline{41.72} & \cellcolor{lightgray}50.99 & \cellcolor{lightgray} 28.19 & 27.87 & \underline{23.70} & 27.93 & 23.74 & 23.88 & 26.72 \\ 
\midrule 
\multirow{2}{*}{Metric} & \multicolumn{7}{c}{OpenStack Raw:15.36 Formatted:18.23} & \multicolumn{7}{c}{HPC Raw:18.27 Formatted:32.84} \\ 
\cmidrule(lr){2-8} \cmidrule(lr){9-15} 
& Drain & AEL & IPLoM & LFA & LogCluster & Brain & GT & Drain & AEL & IPLoM & LFA & LogCluster & Brain & GT \\ 
\midrule 
GA (\%) & 75.2 & 74.3 & 36.5 & 66.9 & 68.5 & 100.0 & \cellcolor{lightgray} 100.0 & 79.3 & 74.8 & 78.7 & 72.8 & 72.5 & 80.0 & \cellcolor{lightgray} 100.0 \\ 
FGA (\%) & 00.7 & 68.2 & 02.8 & 60.0 & 00.1 & 100.0 & \cellcolor{lightgray} 100.0 & 30.9 & 20.1 & 24.0 & 34.7 & 04.6 & 44.5 & \cellcolor{lightgray} 100.0 \\ 
PA (\%) & 02.9 & 02.9 & 00.0 & 00.8 & 01.0 & 29.2 & \cellcolor{lightgray} 100.0 & 72.1 & 74.1 & 64.0 & 69.2 & 68.2 & 66.3 & \cellcolor{lightgray} 100.0 \\ 
RTA (\%) & 14.6 & 14.6 & 00.0 & 06.2 & 04.2 & 29.2 & \cellcolor{lightgray} 100.0 & 40.5 & 48.6 & 35.1 & 29.7 & 39.2 & 40.5 & \cellcolor{lightgray} 100.0 \\ 
PTA (\%) & 00.1 & 18.9 & 00.0 & 09.4 & 00.0 & 29.2 & \cellcolor{lightgray} 100.0 & 09.3 & 07.9 & 07.6 & 17.2 & 01.9 & 21.3 & \cellcolor{lightgray} 100.0 \\ 
FTA (\%) & 00.2 & 16.5 & 00.0 & 07.5 & 00.0 & 29.2 & \cellcolor{lightgray} 100.0 & 15.2 & 13.6 & 12.5 & 21.8 & 03.6 & 21.3 & \cellcolor{lightgray} 100.0 \\ 
CR & \underline{16.81} & 18.70 & 18.75 & 18.81 & 18.12 & 20.69 & \cellcolor{lightgray} 20.78 & 32.40 & 32.72 & 32.38 & \underline{32.37} & \cellcolor{lightgray} 33.12 & 32.96 & 33.27 \\ 
\midrule 
\multirow{2}{*}{Metric} & \multicolumn{7}{c}{Mac Raw:23.24 Formatted:32.26} & \multicolumn{7}{c}{OpenSSH Raw:19.94 Formatted:48.41} \\ 
\cmidrule(lr){2-8} \cmidrule(lr){9-15} 
& Drain & AEL & IPLoM & LFA & LogCluster & Brain & GT & Drain & AEL & IPLoM & LFA & LogCluster & Brain & GT \\ 
\midrule 
GA (\%) & 76.1 & 79.7 & 65.3 & 59.2 & 46.2 & 83.4 & \cellcolor{lightgray} 100.0 & 70.7 & 70.5 & 77.9 & 15.5 & 21.6 & 66.3 & \cellcolor{lightgray} 100.0 \\ 
FGA (\%) & 22.9 & 79.3 & 64.5 & 64.6 & 06.8 & 75.4 & \cellcolor{lightgray} 100.0 & 87.2 & 68.9 & 89.7 & 50.6 & 00.6 & 73.7 & \cellcolor{lightgray} 100.0 \\ 
PA (\%) & 35.7 & 24.5 & 07.3 & 22.7 & 21.3 & 32.5 & \cellcolor{lightgray} 100.0 & 58.6 & 36.4 & 19.4 & 05.4 & 18.8 & 48.1 & \cellcolor{lightgray} 100.0 \\ 
RTA (\%) & 25.1 & 20.1 & 26.7 & 08.0 & 14.1 & 33.5 & \cellcolor{lightgray} 100.0 & 50.0 & 39.5 & 21.1 & 15.8 & 31.6 & 39.5 & \cellcolor{lightgray} 100.0 \\ 
PTA (\%) & 04.0 & 20.9 & 22.0 & 09.1 & 01.0 & 33.5 & \cellcolor{lightgray} 100.0 & 47.5 & 28.8 & 20.0 & 14.6 & 00.3 & 34.5 & \cellcolor{lightgray} 100.0 \\ 
FTA (\%) & 06.9 & 20.5 & 24.1 & 08.5 & 01.9 & 29.4 & \cellcolor{lightgray} 100.0 & 48.7 & 33.3 & 20.5 & 15.2 & 00.5 & 34.5 & \cellcolor{lightgray} 100.0 \\ 
CR & \underline{37.38} & \cellcolor{lightgray}43.97 & 37.81 & 37.60 & 40.60 & 41.94 & 41.65 & \cellcolor{lightgray}75.67 & 74.89 & 74.58 & 73.27 & \underline{58.62} & 59.77 & 75.64 \\ 
\midrule 
\multirow{2}{*}{Metric} & \multicolumn{7}{c}{BGL Raw:21.27 Formatted:34.19} & \multicolumn{7}{c}{HDFS Raw:15.26 Formatted:18.88} \\ 
\cmidrule(lr){2-8} \cmidrule(lr){9-15} 
& Drain & AEL & IPLoM & LFA & LogCluster & Brain & GT & Drain & AEL & IPLoM & LFA & LogCluster & Brain & GT \\ 
\midrule 
GA (\%) & 91.9 & 91.5 & 89.6 & 73.3 & 75.6 & 92.0 & \cellcolor{lightgray} 100.0 & 99.9 & 99.9 & 90.9 & 74.8 & 55.8 & 96.0 & \cellcolor{lightgray} 100.0 \\ 
FGA (\%) & 62.4 & 58.7 & 69.2 & 43.2 & 00.3 & 69.5 & \cellcolor{lightgray} 100.0 & 93.5 & 76.4 & 00.5 & 81.8 & 00.0 & 66.7 & \cellcolor{lightgray} 100.0 \\ 
PA (\%) & 40.7 & 40.6 & 33.8 & 22.8 & 30.6 & 39.5 & \cellcolor{lightgray} 100.0 & 62.1 & 62.1 & 29.0 & 15.3 & 40.3 & 92.9 & \cellcolor{lightgray} 100.0 \\ 
RTA (\%) & 22.2 & 18.4 & 12.5 & 05.6 & 14.4 & 23.7 & \cellcolor{lightgray} 100.0 & 60.9 & 54.3 & 26.1 & 04.3 & 06.5 & 66.7 & \cellcolor{lightgray} 100.0 \\ 
PTA (\%) & 09.3 & 14.9 & 12.9 & 05.0 & 00.0 & 18.4 & \cellcolor{lightgray} 100.0 & 60.9 & 58.1 & 00.0 & 04.8 & 00.0 & 66.7 & \cellcolor{lightgray} 100.0 \\ 
FTA (\%) & 17.1 & 16.5 & 12.7 & 05.3 & 00.1 & 18.4 & \cellcolor{lightgray} 100.0 & 60.9 & 56.2 & 00.1 & 04.5 & 00.0 & 66.7 & \cellcolor{lightgray} 100.0 \\ 
CR & \cellcolor{lightgray}35.20 & 34.88 & 34.80 & 35.06 & 34.39 & \underline{34.03} & 34.94 & 19.49 & \cellcolor{lightgray} 19.59 & 18.98 & 19.30 & \underline{18.82} & 19.24 & 19.47 \\ 
\bottomrule 
\end{tabular}
\label{cr_pc} 
\end{adjustbox} 
\end{table*}

\subsection{Answer to Q1}

First, we confirm that parsing is beneficial. As shown in Table~\ref{cr_pc}, applying parsing improves the compression ratio (CR) over raw logs by 9.4\% to 194.4\% across all datasets.

However, the central question is whether \textit{higher accuracy} leads to \textit{better compression}. To investigate this, we measured the Spearman correlation between six parsing accuracy metrics and the CR. The results in Figure~\ref{spearman_fig}: out of 84 correlations, only 12 have an absolute value greater than 0.7. No single metric shows a stable relationship with CR across datasets. When aggregated (Figure~\ref{spearman_all}), the correlation remains weak, with the highest coefficient being a mere 0.280.

These results strongly suggest a \textbf{misalignment between the objectives of current log parsers (i.e., template accuracy) and the actual requirements of log compression}.

\begin{tcolorbox}[colback=gray!10, colframe=black, sharp corners, boxrule=0.8pt]
\textbf{Answer to Q1:} Log parsing enhances the compression ratio, but higher parsing accuracy does not reliably lead to a better compression ratio.
\end{tcolorbox}

\subsection{Answer to Q2}
The weak correlation found in Q1 naturally leads to Q2: if not parsing accuracy, what structural properties of a processed log make it highly compressible? Our analysis, which deconstructed the parsing pipeline and investigated the underlying mechanisms, reveals one dominant principle: \textbf{effective pattern-based grouping}.

\paragraph{Formatting, as a Basic Pattern-based Grouping, Dominates Compression Gains.}
Most parsers operate in two stages: log formatting and template extraction. We found that the initial formatting stage, a basic pattern-based grouping of structured fields like \texttt{<Time>} and \texttt{<Level>}, accounts for the vast majority of compression gains. For example, on the OpenStack dataset, formatting alone achieves a CR of 18.23. Subsequent template extraction by even the best-performing parser, LFA, only incrementally increases this to 18.81. This phenomenon is prevalent in datasets like HPC, and BGL, where template extraction provides minimal or even slightly negative improvements (due to template indexing overhead). 
For the other datasets, the formatting stage is also responsible for the majority of the compression ratio gains from log parsing, typically contributing more than half of the total improvement. Taking the Zookeeper dataset as an example, formatting alone raised the compression ratio from 27.55 (raw) to 61.69, a gain of over 34, while further event parsing only added about 15. Likewise, in the Hadoop dataset, formatting contributed to 2/3 of the overall improvement.

\begin{table*}[t]
\centering
\caption{The Compression Ratio (CR) of Specialized Log Compressors and General-Purpose Compressors. The best result is in \textbf{bold}.}
\label{tab:compression_ratio_expanded}

\begin{tabular}{lcccccccc}
\toprule
Dataset & gzip & bzip2 & lzma & LogReducer & LogShrink & Denum & DeLog & DeLog-L \\
\midrule
Android     & 7.515   & 12.740  & 19.399  & 20.776 & 21.857 & \textbf{32.494} & 30.354 & 28.634 \\
Apache      & 20.872  & 31.589  & 30.754  & 43.028 & 55.940 & 58.517 & \textbf{59.648} & 49.872 \\
BGL         & 11.808  & 15.463  & 22.053  & 38.600 & 42.385 & 41.804 & \textbf{45.682} & 36.106 \\
Hadoop      & 19.207  & 32.358  & 37.592  & 52.830 & 60.091 & 78.546 & \textbf{79.626} & 67.660 \\
HDFS        & 10.144  & 14.048  & 15.615  & 22.634 & 27.319 & 25.670 & \textbf{29.013} & 28.386 \\
HealthApp   & 10.283  & 13.823  & 15.830  & 31.694 & 39.072 & 44.472 & \textbf{55.934} & 31.921 \\
HPC         & 10.215  & 12.741  & 19.787  & 32.070 & 35.878 & 45.275 & \textbf{46.842} & 26.946 \\
Linux       & 10.133  & 14.723  & 19.984  & 25.213 & 29.252 & 30.449 & \textbf{30.626} & 21.149 \\
Mac         & 11.385  & 18.111  & 23.878  & 35.251 & 39.860 & 40.789 & \textbf{43.687} & 40.923 \\
OpenSSH     & 16.042  & 22.858  & 22.266  & 86.699 & 103.175 & 101.654 & \textbf{106.012} & 60.313 \\
OpenStack   & 11.387  & 15.237  & 15.654  & 16.701 & 22.157 & 22.238 & \textbf{23.754} & 21.783 \\
Proxifier   & 14.838  & 23.673  & 21.086  & 25.501 & 27.029 & 27.288 & \textbf{30.784} & 28.528 \\
Spark       & 16.049  & 26.461  & 23.131  & 59.470 & 59.739 & 59.470 & \textbf{61.788} & 48.701 \\
Thunderbird & 15.755  & 25.363  & 28.536  & 49.185 & 48.434 & 63.824 & \textbf{68.647} & 61.984 \\
Windows     & 16.811  & 67.489  & 217.846 & 342.975 & 456.301 & 481.350 & \textbf{541.569} & 519.578 \\
Zookeeper   & 22.695  & 36.164  & 30.020  & 94.562 & 116.981 & 135.251 & \textbf{154.926} & 47.373 \\
\midrule
Average     & 13.370  & 23.738  & 34.499  & 60.928 & 74.091 & 80.568 & \textbf{88.056} & 69.991 \\
\bottomrule
\end{tabular}%

\end{table*}

\paragraph{Effective Template Extraction is also a Form of Refined Pattern-based Grouping.}
Following our first insight, we investigate the question: When does event parsing actually contribute or not to compression? We begin by examining cases where it provides minimal benefit to formatted logs. In the BGL dataset, for instance, the maximum improvement in compression ratio from event parsing is less than one point. We found that BGL logs are structurally simple: excluding the header, out of over 4.6 million log lines, approximately two million contain no variables, and another 1.7 million follow the pattern generating core.XXXX, where XXXX represents a randomly generated number ranging from 0 to 100,000.
According to the prevailing theory in log compression research, extracting such variable converting "generating core.123" into a template ID should eliminate significant redundancy. However, our results show that this process yields a negligible improvement in compression ratio.
To understand what constitutes truly effective variable extraction, we turn our attention to the Apache dataset. Here, LogCluster initially achieves a conspicuously lower compression ratio. The root cause is its failure to correctly extract a numerical variable, the Child ID, which is of a similar type to the variables in BGL. This failure creats thousands of distinct templates like "Child 1234" and "Child 5233". Why does this failure have such a dramatic impact on compression? Because unlike the random numbers in BGL, the Child ID in Apache logs is highly patterned—it typically increments sequentially by one. After we assisted LogCluster in correctly extracting the Child ID, its compression ratio rose to a level on par with the other parsers.

\paragraph{Optimal Pattern-based Grouping Can Outperform Accurate Template Extraction.}
The results show that the highest compression ratio is often achieved not by the most accurate parser, but by the one that creates the most homogeneous, compressible groups, even if it means sacrificing formal parsing accuracy.
On the Proxifier dataset, \textit{IPLoM} achieves a CR of 26.4, higher than the Ground Truth's CR of 24.5. The Ground Truth correctly places all domain names into a single group under one wildcard. IPLoM, however, incorrectly splits this group into two: one for the \textit{LAN domain pattern} (e.g., \texttt{"dshytnh:80"}) and one for the \textit{remote domain pattern} (e.g., \texttt{"mtalk.google.com:5228"}). This ``inaccurate'' but more refined grouping was superior because it created two more internally homogeneous, compressible, data streams. In essence, sacrificing formal accuracy to create more refined, pattern-homogeneous groups directly improved the compression ratio. 

\paragraph{Conclusion.}
These findings collectively demonstrate that pursuing parsing accuracy is a flawed strategy for log compression. The true driver of efficiency is the creation of \textbf{homogeneous groups of words that share a common, compressible pattern}.  \textit{DeLog}.

\begin{tcolorbox}[colback=gray!10, colframe=black, sharp corners, boxrule=0.8pt]
\textbf{Answer to Q2:} A log representation is compression-friendly when it \textbf{achieves pattern-based grouping}, as this creates homogeneous data streams amenable to specialized encoding. This is more critical than achieving perfect parsing accuracy.
\end{tcolorbox}

\begin{table*}[t]
\centering
\caption{The Compression Speed (MB/s) of Specialized Log Compressors and General-Purpose Compressors. The best result is in \textbf{bold}.}
\label{tab:compression_speed_expanded}

\begin{tabular}{lcccccccc}
\toprule
Dataset & gzip & bzip2 & lzma & LogReducer & LogShrink & Denum & DeLog & DeLog-L \\
\hline
Android     & \textbf{118.382} & 34.162 & 17.675 & 3.096  & 0.448  & 15.427 & 25.232 & 22.423  \\
Apache      & \textbf{25.975}  & 5.335  & 5.877  & 0.665  & 0.292  & 3.881  & 7.842  & 6.325   \\
BGL         & \textbf{137.964} & 34.158 & 12.234 & 12.020 & 1.623  & 16.189 & 27.890 & 22.341  \\
Hadoop      & \textbf{138.091} & 20.306 & 16.914 & 2.976  & 0.508  & 17.449 & 32.393 & 29.784  \\
HDFS        & \textbf{146.022} & 25.650 & 11.373 & 14.280 & 1.812  & 17.444 & 31.232 & 29.932  \\
HealthApp   & \textbf{78.193}  & 22.717 & 7.002  & 0.681  & 1.220  & 10.550 & 19.523 & 14.032  \\
HPC         & \textbf{71.300}  & 24.759 & 9.402  & 3.850  & 1.430  & 10.617 & 20.081 & 16.849  \\
Linux       & \textbf{14.326}  & 5.492  & 3.026  & 0.159  & 0.077  & 2.394  & 3.893  & 3.423   \\
Mac         & \textbf{49.912}  & 9.354  & 6.399  & 0.620  & 0.210  & 5.018  & 10.423 & 10.042  \\
OpenSSH     & \textbf{123.861} & 23.819 & 11.661 & 6.656  & 1.365  & 14.919 & 30.683 & 28.823  \\
OpenStack   & \textbf{112.333} & 13.143 & 7.938  & 5.516  & 1.142  & 9.735  & 14.978 & 12.734  \\
Proxifier   & \textbf{14.849}  & 4.924  & 4.295  & 0.356  & 0.060  & 2.679  & 4.231  & 4.534   \\
Spark       & \textbf{139.263} & 28.948 & 15.804 & 12.292 & 1.320  & 16.828 & 28.342 & 26.442  \\
Thunderbird & \textbf{181.358} & 29.166 & 20.265 & 8.435  & 1.321  & 13.460 & 29.181 & 28.712  \\
Windows     & \textbf{243.221} & 16.929 & 50.948 & 13.138 & 1.931  & 20.617 & 46.238 & 46.374  \\
Zookeeper   & \textbf{43.249}  & 6.797  & 5.385  & 1.331  & 0.432  & 5.466  & 10.003 & 8.321   \\ 
\hline
Average     & \textbf{96.144} & 18.428 & 12.887 & 5.379  & 0.950  & 11.417 & 21.394 & 19.435  \\ 
\bottomrule
\end{tabular}%

\end{table*}

\section{Extended Experimental Results}\label{empirical}

\subsection{Experimental Setting}
The experimental environment, including the hardware and software platforms, is identical to that detailed in our main paper.

To ensure a fair comparison, all general-purpose compressors (\texttt{gzip}, \texttt{bzip2}, and \texttt{lzma}) were configured to use their highest compression level (e.g., \texttt{-9} for \texttt{gzip}). Furthermore, the input log files were partitioned into 100K lines blocks, and these blocks were subsequently compressed in parallel using 4 threads. This block-based, multi-threaded approach was chosen to maintain consistency with the processing model of the specialized log compressors, thus providing a comparable performance baseline.

\paragraph{Compression Ratio Comparison with General-Purpose Compressors}
Table \ref{tab:compression_ratio_expanded} extends the analysis from the main paper by comparing the compression ratio (CR) of specialized log compressors against widely-used general-purpose compressors, namely \texttt{gzip}, \texttt{bzip2}, and \texttt{lzma}. The results unequivocally demonstrate the significant advantage of a domain-specific approach for log data. DeLog achieves an average CR of 88.056, which is more than 2.5 times higher than \texttt{lzma} (34.499), the most effective general-purpose compressor in this comparison. This performance gap underscores the limitations of general-purpose algorithms, which treat logs as generic text and fail to exploit the highly structured and repetitive nature of log templates and parameters. Even the baseline specialized compressors like LogReducer and Denum consistently outperform \texttt{gzip}, \texttt{bzip2}, and \texttt{lzma}, validating the core motivation for developing dedicated log compression solutions to maximize storage savings.

\paragraph{Compression Speed Comparison with General-Purpose Compressors}
Table \ref{tab:compression_speed_expanded} provides a comparative analysis of compression speed, revealing a classic trade-off between speed and compression ratio. As the data shows, \texttt{gzip} achieves the highest compression speed across nearly all datasets, with an average speed of 96.1 MB/s. This exceptional speed is a key reason for its widespread use in scenarios where low latency is critical, such as on-the-fly data transmission. However, as established in Table \ref{tab:compression_ratio_expanded}, this speed comes at the cost of a significantly lower compression ratio. In contrast, DeLog (21.4 MB/s) and DeLog-L (19.4 MB/s) offer a more balanced profile. While not as fast as \texttt{gzip}, they are substantially faster than other high-ratio compressors like \texttt{lzma} (12.9 MB/s) and \texttt{bzip2} (18.4 MB/s), and significantly outperform their direct competitors in the specialized domain. This positions DeLog as a strong candidate for scenarios requiring both high compression efficiency and robust throughput, occupying a ``sweet spot'' that general-purpose tools cannot fill.

\paragraph{Decompression Speed Comparison with General-Purpose Compressors}
Table \ref{tab:decompression_speed_expanded} details the decompression performance of all methods, a critical metric for log analysis and query workloads. The results highlight the exceptional design of DeLog-L, which achieves the highest decompression speed on every single dataset, averaging 177.6 MB/s. This performance surpasses not only other specialized compressors but also fast general-purpose tools like \texttt{gzip} (166.6 MB/s on average). The standard DeLog is also highly competitive, performing on par with leading specialized tools and faster than many general-purpose ones. This superior decompression speed is attributable to DeLog's structured format, which enables efficient and direct data reconstruction. The analysis confirms that while general-purpose compressors offer a range of performance profiles, DeLog and its variants provide a solution that uniquely combines state-of-the-art compression ratios with best-in-class decompression speeds, making it ideally suited for modern log management systems.

\begin{table*}[ht]
\centering
\caption{The Decompression Speed (MB/s) of Specialized Log Compressors and General-Purpose Compressors. The best result is in \textbf{bold}.}
\label{tab:decompression_speed_expanded}
\begin{tabular}{lcccccccc}
\toprule
Dataset & gzip & bzip2 & lzma & LogReducer & LogShrink & Denum & DeLog & DeLog-L \\
\hline
Android     & 204.995 & 109.704 & 186.546 & 1.084  & 0.162  & 135.982 & 123.230 & \textbf{221.034} \\
Apache      & 33.834  & 24.242  & 33.612  & 1.516  & 0.532  & 24.724  & 36.434  & \textbf{59.342}  \\
BGL         & 255.416 & 135.767 & 215.952 & 7.482  & 0.872  & 196.472 & 134.193 & \textbf{276.694} \\
Hadoop      & 205.961 & 97.823  & 195.559 & 3.055  & 0.189  & 66.891  & 103.101 & \textbf{223.342} \\
HDFS        & 254.291 & 112.512 & 213.144 & 8.674  & 0.881  & 172.330 & 181.481 & \textbf{294.743} \\
HealthApp   & 125.632 & 62.419  & 105.974 & 2.071  & Error  & 28.601  & 69.543  & \textbf{134.534} \\
HPC         & 98.780  & 66.755  & 83.820  & 2.170  & 1.107  & 76.910  & 78.203  & \textbf{140.324} \\
Linux       & 15.338  & 12.094  & 13.209  & 0.233  & 0.096  & 7.770   & 22.004  & \textbf{29.231}  \\
Mac         & 79.649  & 39.054  & 74.091  & 0.493  & 0.123  & 26.991  & 45.943  & \textbf{92.439}  \\
OpenSSH     & 177.018 & 102.718 & 144.722 & 4.404  & 0.730  & 173.240 & 143.231 & \textbf{243.234} \\
OpenStack   & 203.104 & 69.388  & 144.861 & 6.546  & 0.866  & 67.512  & 89.782  & \textbf{153.423} \\
Proxifier   & 15.359  & 12.622  & 15.108  & 0.083  & 0.071  & 7.879   & 20.799  & \textbf{43.433}  \\
Spark       & 191.603 & 113.142 & 163.529 & 3.781  & 0.520  & 197.253 & 185.762 & \textbf{275.839} \\
Thunderbird & 273.692 & 144.131 & 252.886 & 3.556  & 0.437  & 112.234 & 156.234 & \textbf{327.481} \\
Windows     & 370.157 & 144.748 & 414.001 & 11.487 & 0.823  & 233.171 & 282.328 & \textbf{548.323} \\
Zookeeper   & 60.605  & 35.427  & 56.026  & 0.456  & 0.733  & 30.532  & 38.932  & \textbf{78.343}  \\ 
\hline
Average     & 166.584 & 86.447  & 147.253 & 3.503  & 0.476  & 97.406  & 98.200  & \textbf{177.605} \\ 
\bottomrule
\end{tabular}%
\end{table*}

\section*{Revisiting the Evaluation of Log Parsers.}

\textbf{Revisiting the Evaluation of Log Parsers.} In reproducing the experiments of existing log parsers, we discovered significant issues within the LogPAI evaluation framework that lead to inaccurate accuracy measurements. Our reported accuracies therefore differ from those in many previous studies, and we contend that our figures represent a more precise assessment. The discrepancies stem from two primary sources of error in the original LogPAI implementation, which we have corrected:

\begin{enumerate}
    \item \textbf{Incorrect Ground Truth Preprocessing:} The evaluation script is intended to preprocess ground truth logs by replacing known variable patterns with a generic \texttt{<*>} placeholder. However, a critical implementation bug exists for some parsers where the correct code, \texttt{line = re.sub(currentRex, "<*>", line)}, was mistakenly written as \texttt{line = re.sub(currentRex, "", line)}. This error, which replaces variables with an empty string instead of the placeholder, fundamentally alters the log structure and invalidates the subsequent accuracy comparison.

    \item \textbf{Inconsistent Placeholder Representation:} The framework fails to correctly evaluate parsers that use non-standard placeholders. For instance, LogCluster generates valid variable tags like \texttt{*\{1.1\}}, which encode variable attributes. The LogPAI evaluation script, hard-coded to expect only \texttt{<*>}, incorrectly marks these correct results as parsing failures.
\end{enumerate}

By rectifying these issues in our evaluation harness, we ensure a fair and accurate comparison across all parsers. The impact of these corrections is substantial, leading to accuracy differences of tens of percentage points compared to the original, uncorrected framework. This suggests that the baseline accuracies reported in numerous log parsing papers may require re-evaluation.

\end{document}